\documentclass[aps,prd,amsmath,twocolumn,showpacs]{revtex4}

\usepackage{epsfig}
\usepackage{epstopdf}
\usepackage{graphics}
\usepackage{latexsym}
\usepackage{tabularx}
\usepackage{amsmath}
\usepackage{amssymb}
\usepackage{array,booktabs,makecell}
\usepackage{rotating}
\usepackage{subfigure}
\usepackage{bm}
\usepackage{fixmath}
\usepackage{xcolor}

\usepackage[colorlinks = true,
linkcolor = black,
urlcolor  = blue,
citecolor = black,
anchorcolor = blue]{hyperref}

{ 

\makeatletter
\def\mathcolor#1#{\@mathcolor{#1}}
\def\@mathcolor#1#2#3{%
  \protect\leavevmode
  \begingroup
    \color#1{#2}#3%
  \endgroup
}
\makeatother

\begin{document}

\title{Determination of generalized parton distributions through a simultaneous analysis of axial form factor and wide-angle Compton scattering data}

\author{Hadi Hashamipour$^{a}$}
\email{h\_hashamipour@sbu.ac.ir}

\author{Muhammad Goharipour$^{b}$}
\email{muhammad.goharipour@ipm.ir}

\author{Siamak S. Gousheh$^{a}$}
\email{ss-gousheh@sbu.ac.ir}
\affiliation {
$^{a}$Department of Physics, Shahid Beheshti University, G.C., Evin, Tehran 19839, Iran  \\
$^{b}$School of Particles and Accelerators, Institute for Research in Fundamental Sciences (IPM), PO Box 19568-36681, Tehran, Iran}

\date{\today}

\begin{abstract}

In this work, we present a new set of unpolarized ($ H $) and polarized ($\widetilde{H}$) generalized parton distributions (GPDs) that have been determined using a simultaneous $ \chi^2 $ analysis of the nucleon axial form factor (AFF) and wide-angle Compton scattering (WACS) experimental data at the next-to-leading order (NLO) accuracy in QCD. We explore various Ansatzes presented in the literature for GPDs, which use forward parton distributions as input, and choose the ones most suited to our analysis. The experimental data included in our analysis cover a wide range of the squared transverse momentum, which is $ 0.025 < -t < 6.46 $ GeV$ ^2 $. We show that the WACS data affect significantly the large $-t$ behavior of $\widetilde{H}$. 
The polarized GPDs obtained from the simultaneous analysis of AFF and WACS data  differ considerably from the corresponding ones obtained by analyzing AFF and WACS separately, and have less uncertainties. We show that the theoretical predictions obtained using our GPDs are in good agreement with the analyzed AFF and WACS data for the entire range of $ -t $ studied. Finally, we obtain the impact parameter dependent parton distributions, both in an unpolarized and in a transversely polarized proton, and present them as tomography plots.

\end{abstract}


%
\maketitle

\section{Introduction}\label{sec:one} 
The factorization theorem has been very successful in describing perturbative quantum chromodynamics (QCD) processes, considering them as being composed of a soft nonperturbative and a hard parton level (perturbatively calculable) part. Many processes that are used to understand the structure of hadrons such as the deep inelastic scattering (DIS), deeply virtual Compton scattering (DVCS), deeply virtual meson production (DVMP), and wide-angle Compton scattering (WACS) can be studied using the factorization theorem and perturbative QCD analysis. It is well known that the nonperturbative part can be described using the language of parton distribution functions (PDFs)~\cite{Pumplin:2002vw,Jimenez-Delgado:2014twa,Harland-Lang:2014zoa,Ball:2014uwa,Butterworth:2015oua,Dulat:2015mca,Alekhin:2017kpj,Ball:2017nwa,AbdulKhalek:2019bux,AbdulKhalek:2019ihb,Radyushkin:2019mye} and also polarized PDFs (PPDFs)~\cite{deFlorian:2009vb,Blumlein:2010rn,Leader:2010rb,deFlorian:2014yva,Nocera:2014gqa,Jimenez-Delgado:2014xza,Sato:2016tuz,Shahri:2016uzl,Khanpour:2017cha,Ethier:2017zbq,Salajegheh:2018hfs}. In fact, these nonperturbative objects, which are usually extracted from the experimental data by the well-established means of global analysis, play a crucial role in all calculations of high-energy processes with initial hadrons. However, the structure of the nucleon in both the unpolarized and polarized cases can be investigated in more detail using generalized parton distributions (GPDs)~\cite{Goeke:2001tz,Diehl:2003ny,Polyakov:2002yz,Freund:2002qf,Scopetta:2003et,Belitsky:2005qn,Boffi:2007yc,Guzey:2005ec,Guzey:2006xi,Goeke:2008jz,Hagler:2007xi,Alexandrou:2013joa,Kumericki:2007sa,Guidal:2013rya,Kumericki:2016ehc,Khanpour:2017slc,Kroll:2017hym,Kaur:2018ewq,Moutarde:2018kwr,Sun:2018ldr,deTeramond:2018ecg,Duplancic:2018bum,Siddikov:2019ahb,Kriesten:2019jep,Hashamipour:2019pgy,Kroll:2020jat} which are directly related to amplitudes of physical processes in Bjorken kinematics~\cite{Ji:1998xh,Collins:1998be}.

GPDs display the characteristic properties to present a three-dimensional (3D) description of hadrons since they provide quantitative information on both the longitudinal and transverse distributions of partons
inside the nucleon, and also their intrinsic and orbital angular momenta. Indeed, they can be easily reduced to PDFs, form factors (FFs), charge distributions, magnetization density, and
gravitational form factors. Generally, GPDs are functions of three variables $ x $, $ \xi $ and $ t $. The variables $ x $ and $ t $ are the fraction of momentum carried by the active quark and the square of the momentum transfer in the process, respectively, while $ \xi $ gives the longitudinal momentum transfer.
It was recognized from the beginning that the exclusive scattering processes like DVCS~\cite{Kroll:1995pv,Ji:1996nm,Guichon:1998xv,Belitsky:2001ns,Belitsky:2012ch,Benali:2020vma} or DVMP~\cite{Mueller:1998fv,Collins:1996fb,Favart:2015umi,Brooks:2018uqk} are an excellent way to probe GPDs. However, because of the poorly known wave functions of the produced mesons as well as the sizable higher twist contributions, additional channels are needed to get further information on GPDs. It is well known now that other exclusive processes such as the time-like Compton scattering~\cite{Berger:2001xd,Moutarde:2013qs,Boer:2015fwa}, $ \rho $-meson photoproduction~\cite{Diehl:1998pd,Mankiewicz:1999tt,Anikin:2009bf}, heavy vector meson production~\cite{Ivanov:2004vd}, double deeply virtual Compton scattering~\cite{Guidal:2002kt,Belitsky:2002tf}, exclusive pion- or photon-induced lepton pair production~\cite{Muller:2012yq,Sawada:2016mao}, two particles~\cite{Boussarie:2016qop,Pedrak:2017cpp} and neutrino induced exclusive reactions~\cite{Kopeliovich:2012dr,Pire:2017tvv,Pire:2017lfj}, as well as a few other channels~\cite{Kofler:2014yka,Accardi:2012qut}, can also provide information on GPDs.

Although the first Mellin moments of GPDs in special cases can be determined from the lattice QCD calculations~\cite{Hagler:2007xi,Alexandrou:2013joa} and also there are early studies of GPDs using various dynamical models of the nucleon structure~\cite{Khanpour:2017slc}, the well-established method to extract GPDs is analyzing the related experimental data through a QCD fit~\cite{Kumericki:2016ehc}, same as for the PDFs and PPDFs. To this end, there have been various models~\cite{Pasquini:2005dk,Pasquini:2006dv,Dahiya:2007mt,Frederico:2009fk,Mukherjee:2013yf,Maji:2015vsa} and
parameterizations~\cite{Goldstein:2010gu,Goldstein:2013gra,Sharma:2016cnf} for GPDs during the last two decades. In the early analyses of GPDs, the experimental data from DVCS and DVMP were mostly used. In fact, there are valuable data provided by the H1, ZEUS and HERMES Collaborations at DESY, in addition to some measurements by the CLAS and Hall A Collaborations at JLab which cover a wide kinematical region~\cite{Kumericki:2016ehc}. Note that the HERMES, CLAS and Hall A measurements were performed with a fixed proton target. Fortunately, some forthcoming experiments are also being done at upgraded the JLab~\cite{Kubarovsky:2011zz,Armstrong:2017wfw}, COMPASS~\cite{dHose:2004usi,Silva:2013dta,Kouznetsov:2016vvo} and J-PARC~\cite{Sawada:2016mao,Kroll:2016kvd} which can provide further constraints on GPDs. Moreover, there are planned experiments at the electron ion collider (EIC)~\cite{Accardi:2012qut} and large hadron electron collider (LHeC)~\cite{AbelleiraFernandez:2012cc}, where the measurements of exclusive processes are among the main goals of their experimental programs.

As mentioned, FFs, whether the electric and magnetic form factors or those associated with the energy-momentum tensor, can be obtained from GPDs~\cite{Diehl:2004cx,Diehl:2007uc,Diehl:2013xca} through the
so-called Ji's sum rule~\cite{Ji:1996nm,Ji:1996ek}. In this regard, the nucleon axial form factor
(AFF), that describe spin content of the nucleon, is also related to polarized GPDs. There are various approaches to extract AFF including lattice QCD calculations and neural networks (see Ref.~\cite{Hashamipour:2019pgy} and references therein).  It can also be obtained from the eigenstates of a light-front effective Hamiltonian in the leading Fock representation~\cite{Xu:2019xhk}. One can refer to Refs.~\cite{Bernard:2001rs,Schindler:2006jq} to get a review of AFF experimental data. In our previous work~\cite{Hashamipour:2019pgy}, we used a practical Ansatz suggested by Diehl, Feldmann, Jakob, and Kroll (DFJK)~\cite{Diehl:2004cx,Diehl:2007uc,Diehl:2013xca}, which relates the predetermined (polarized) PDFs as input to (polarized) GPDs, to extract the polarized GPDs for quarks ($\widetilde{H}$) through a standard $\chi^2$ analysis of the nucleon AFF data. We showed that some parameters of the model should be readjusted to obtain better consistency between the theoretical predictions and experimental data. In this work, we are going to continue our studies in this area by determining GPDs using a simultaneous analysis of AFF and WACS data to investigate the impact of latter one on the extracted GPDs compared to those obtained by analyzing AFF data solely. Actually, our motivation comes from the recent Kroll's work~\cite{Kroll:2017hym}, where it has been shown that the WACS data can be used to constrain the large $ -t $ behavior of $\widetilde{H}$. 

\begin{figure}[ht!]
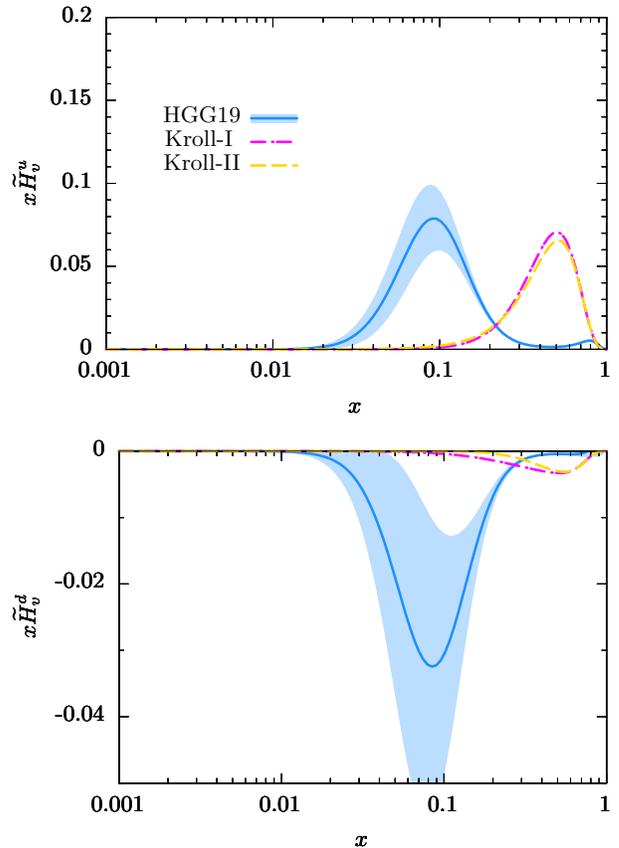
\label{fig:fig1}
\resizebox{!}{0.32\textwidth}
{
\input{./fig-huv.tex}
}
\resizebox{!}{0.32\textwidth}
{
\input{./fig-hdv.tex}
}
\caption{A comparison between the Kroll's results~\cite{Kroll:2017hym} (dashed and dashed-dotted curves) for $\widetilde{H}_v^u$ (up panel) and $\widetilde{H}_v^d$ (bottom panel) valence GPDs at $ t=-4 $ GeV$ ^2 $, and corresponding ones from HGG19 analysis~\cite{Hashamipour:2019pgy} (solid curve).}
\end{figure}

In Fig.~\ref{fig:fig1}, we have compared the results obtained from the WACS data~\cite{Kroll:2017hym} (dashed and dashed-dotted curves) for \textit{up} valence $\widetilde{H}_v^u$ (up panel) and \textit{down} valence $\widetilde{H}_v^d$ (bottom panel) polarized GPDs at $ t=-4 $ GeV$ ^2 $ with our previous work~\cite{Hashamipour:2019pgy} (HGG19) that included only the AFF data (solid curve). As can be seen, there are considerable differences between two approaches. To be more precise, for both valence polarized GPDs, Kroll's results are more inclined to larger $ x $ so that they peaked at $ x\sim 0.5 $, while HGG19 results peaked at $ x\sim 0.1 $. This exactly indicates the impact of WACS data on polarized GPDs $\widetilde{H}$, especially at larger values of $ -t $. Another point which should be noted is that for the case of $\widetilde{H}_v^d$, our previous result has a greater magnitude compared to Kroll's result, while both of them have almost same magnitude for the case of $\widetilde{H}_v^u$. As we shall explain in details later, we showed in our previous work~\cite{Hashamipour:2019pgy} that the final results are not very sensitive to the choice of PPDFs set used, i.e.,   \texttt{DSSV08}~\cite{deFlorian:2009vb} and \texttt{NNPDFpol1.1}~\cite{Nocera:2014gqa}.Therefore, the different PPDFs used in these two works (Kroll used \texttt{DSSV08}~\cite{deFlorian:2009vb}, but we used \texttt{NNPDFpol1.1}~\cite{Nocera:2014gqa}) cannot lead to such differences, and the resolution must lie elsewhere. As we shall show, performing a simultaneous analysis of AFF and WACS data leads to an improved polarized GPDs $\widetilde{H}$ which differs from the corresponding ones obtained by analyzing AFF and WACS data separately.
%
%
%

This paper is organized as follows. In Sec.~\ref{sec:two}, we present the theoretical framework we use in this work to analyze the AFF and WACS data and extract GPDs. To this end, following a brief introduction about GPDs, we first review the physics of the theoretical calculation of the nucleon AFF and WACS cross section. Then, we introduce the DFJK model which is used for calculating (polarized) GPDs using predetermined (polarized) PDFs as input. We discuss also the impact parameter dependent PDFs and nucleon
helicity flip distribution $ E $. In Sec.~\ref{sec:three}, we specify the experimental data included in our analysis and describe our procedure of data selection. Sec.~\ref{sec:four} is devoted to presenting the results obtained using various scenarios for theoretical calculation of WACS cross section and also changing input PDFs. The extracted GPDs from different analyses are compared to final GPDs corresponding to the analysis with the lowest value of $ \chi^2 $. Some comparisons between the theoretical predictions obtained using the final GPDs and the experimental data included in the analysis are also presented. Finally, by calculating the distribution in the transverse plane of valence quarks, both in an unpolarized and in a transversely polarized proton, we present our results for `proton tomography'.
We summarize our results and conclusions in Sec.~\ref{sec:five}.

\section{Theoretical Framework}\label{sec:two} 
As pointed out before, GPDs are nonperturbative objects describing soft dynamics inside hadrons. They are in a sense generalization of both FFs and PDFs. Although GPDs cannot be calculated from perturbative QCD, they can be extracted using the standard $ \chi^2 $ analysis of the related experimental data, thanks to the factorization theorem.
In this section, we are going to review some of their features together with their relation to the nucleon AFF, WACS cross section, and Impact parameter dependent PDFs. We also we discuss various Ansatzes for GPDs suggested by DFJK~\cite{Diehl:2004cx,Diehl:2007uc,Diehl:2013xca}.  Here, just like our previous analysis~\cite{Hashamipour:2019pgy}, we use the convention of Ji~\cite{Ji:1996ek} for GPDs, in which $H$, $E$, $\widetilde{H}$ and $\widetilde{E}$ are defined as~\cite{Belitsky:2005qn,Diehl:2003ny}:
\begin{align}
\label{Eq:1}
 \frac{1}{2}\int \frac{d z^-}{2\pi}\, e^{ix P^+ z^-}
\langle p'|\, \bar{q}(-\frac{1}{2} z)\, \gamma^+ q(\frac{1}{2} z) 
\,|p \rangle \Big|_{\substack{z^+=0,\\\bm{z_\perp=0}}}\nonumber \\
= \frac{1}{2P^+}  \left[
H^q(x,\xi,t)\, \bar{u}(p') \gamma^+ u(p) \right.\nonumber \\
 \left.  + E^q(x,\xi,t)\, \bar{u}(p') 
\frac{i \sigma^{+\alpha} \Delta_\alpha}{2m} u(p)
\, \right] ,
\nonumber \\
\frac{1}{2} \int \frac{d z^-}{2\pi}\, e^{ix P^+ z^-}
\langle p'|\, 
\bar{q}(-\frac{1}{2} z)\, \gamma^+ \gamma_5\, q(\frac{1}{2} z)
\,|p \rangle \Big|_{\substack{z^+=0,\\\bm{z_\perp=0}}}\nonumber \\
= \frac{1}{2P^+} \left[
\widetilde{H}^q(x,\xi,t)\, \bar{u}(p') \gamma^+ \gamma_5 u(p)\right.\nonumber \\
\left. + \widetilde{E}^q(x,\xi,t)\, \bar{u}(p') \frac{\gamma_5 \Delta^+}{2m} u(p)
\, \right],
\end{align}
where $z=\left(z^+,\bm{z_\perp},z^-\right)$. As it is evident from Eq.~(\ref{Eq:1}), GPDs depend on three kinematical variables, $ x $, $ \xi $ and $ t $. The first one is the well-known Bjorken scaling variable $x=\frac{Q^2}{2 p\cdot q}$, with photon virtuality $ Q^2 $, which can be interpreted as the average of momentum fractions of active quarks. The other longitudinal variable $\xi=\frac{p^+-p'^+}{p^++p'^+}$, which is called `skewness', does not appear in PDFs. The last argument is $t=(p'-p)^2=\Delta^2$, i.e. the squared of the momentum transferred to the proton target.

We can express valence GPDs $H_v^q$ of flavor $ q $ in terms of quark GPDs $H^q$ as
\begin{equation}
H_v^q(x,t)= H^q (x,\xi=0,t) + H^q (-x,\xi=0,t),
\label{Eq:2}
\end{equation}
with $ H^q (-x,\xi=0,t)= - H^{\bar q} (x,\xi=0,t) $. An analogous relation holds for the valence GPDs $E_v^q$. The situation is somewhat different for the case of valence polarized GPDs $\widetilde{H}^q_v$ so that we have
\begin{equation}
\label{Eq:3}
\widetilde{H}^q_v(x,t) = \widetilde{H}^q(x,\xi=0,t)-\widetilde{H}^q(-x,\xi=0,t),
\end{equation}
with  $ \widetilde{H}^q(-x,\xi=0,t)= \widetilde{H}^{\bar{q}}(x,\xi=0,t) $.

Since we are going to analyze the nucleon AFF and WACS data to put constraints on GPDs, it is worthwhile to review the relevant theoretical framework. To this aim, we will first describe the important sum rules which relate GPDs and FFs in Subsection \ref{sec:two:A}, with more emphasis on the nucleon AFF. The formulas and relations needed for theoretical calculations of the WACS cross section are given in Subsection \ref{sec:two:B}. In Subsection \ref{sec:two:C}, we introduce our phenomenological framework for modeling GPDs and performing a global analysis of AFF and WACS data. Finally, we introduce the impact parameter dependent PDFs and nucleon helicity flip distribution $ E $ that we use for proton tomography in Subsection~\ref{sec:two:D}.

\subsection{Axial Form Factor}\label{sec:two:A}
There are a certain number of sum rules that relate nucleon FFs to GPDs by exploiting the fact that they are different moments of GPDs~\cite{Diehl:2013xca}. The Dirac and Pauli form factors, $ F_1 $ and $ F_2 $, for example, can be written as follows
\begin{align}
F_i^p= e_u F_i^u + e_d F_i^d + e_s F_i^s, \nonumber \\ 
F_i^n= e_u F_i^d + e_d F_i^u + e_s F_i^s, 
\label{Eq:4}
\end{align}
where $ F_i^q $ for $i=1(2)$ is the contribution of quark flavor $q$ to the Dirac (Pauli) FF of the proton ($F^p$) or neutron ($F^n$), respectively. Note that $ e_q $ is the charge of the quark in units of the positron charge. On the other hand, we can write the flavor form factors $ F_i^q $ in terms of the proton valence GPDs $ H_v $ and $ E_v $ for unpolarized quarks of flavor $ q $ as
\begin{align}
F_1^q(t)= \int_0^1 dx~H_v^q(x,t), \nonumber \\ 
F_2^q(t)= \int_0^1 dx~E_v^q(x,t). 
\label{Eq:5}
\end{align}
It is worth noting that the Lorentz invariance makes the result independent of skewness $\xi$, so we choose zero-skewness GPDs and omit this variable from now on.
 
In analogy with the Dirac and Pauli FFs, the nucleon AFF can be expressed in terms of polarized GPDs as~\cite{Diehl:2013xca}  
\begin{align}
G_A(t)=&\int_0^1 dx \left[\widetilde{H}^u_v(x,t)-\widetilde{H}^d_v(x,t)\right]+\nonumber\\
2&\int_0^1 dx \left[\widetilde{H}^{\bar{u}}(x,t)-\widetilde{H}^{\bar{d}}(x,t)\right].
\label{Eq:6}
\end{align}
As it can be readily seen from Eq.~(\ref{Eq:6}), contrary to Pauli and Dirac FFs, the AFF involves some contributions from the sea quarks. We examined these contributions in our previous study \cite{Hashamipour:2019pgy} and found that they are not significant compared to the valence contributions, but not negligible. It is worth mentioning that, from the conceptual point of view, $\int_{0}^{1} dx~\widetilde{H}^q(x,t)$ is the intrinsic spin contribution of quark $q$ to the spin of proton. Some other moments of GPDs can be related to the matrix elements of energy-momentum tensor.

\subsection{Wide-angle Compton scattering}\label{sec:two:B}
As mentioned in the Introduction, we are going to study the impact of WACS data on the behavior of GPDs, especially at larger values of $ -t $, by analyzing them simultaneously with AFF data.  To this end, in this subsection we briefly review the relations and formulae needed to calculate the WACS amplitudes and cross section. If one considers a regime in which the Mandelstam variables $ s $, $ -t $, and $ -u $ are large compared to the QCD scale parameter $ \Lambda $, the (unpolarized) WACS cross section can be written as~\cite{Huang:2001ej},
\begin{align}
{\label{Eq:8}}
\frac{d\sigma}{dt}=&\frac{1}{32\pi\left(s-m^2\right)^2}\left\{\left|\Phi_1\right|^2+\left|\Phi_2\right|^2+2\left|\Phi_3\right|^2+2\left|\Phi_4\right|^2\right.\nonumber\\
&\left.+\left|\Phi_5\right|^2+\left|\Phi_6\right|^2\right\},
\end{align}  
where $\Phi_1, \dots, \Phi_6$ represent the six independent $\Phi_{\mu'\nu',\mu\nu}$ which denote the center-of-mass-system (c.m.s.) helicity amplitudes. In fact, 16 ($2^4$) amplitudes contribute to the Compton scattering theoretically. However, the parity and time-reversal invariance lead to the following relations among them which reduce the number of independent amplitudes to six~\cite{Kroll:2017hym},
\begin{align}
{\label{Eq:9}}
\Phi_{-\mu'-\nu',-\mu-\nu} = \Phi_{\mu\nu,\mu'\nu'}=\left(-1\right)^{\mu-\nu-\mu'+\nu}\Phi_{\mu'\nu',\mu\nu}.
\end{align}  
Below, we show the convention that we have used ~\cite{Huang:2001ej}, 
\begin{align}
{\label{Eq:10}}
\Phi_1 =& \Phi_{++,++},\quad &\Phi_2 =& \Phi_{--,++},\quad &\Phi_3 =& \Phi_{-+,++},\nonumber\\
\Phi_4 =& \Phi_{+-,++},\quad &\Phi_5 =& \Phi_{-+,-+},\quad &\Phi_6 =& \Phi_{-+,+-}.
\end{align}  

It can be shown that, in the handbag approach, the derivation of the Compton amplitudes is inherently simpler using the light-cone helicity basis. On the other hand, the ordinary photon-proton c.m.s. helicity basis is more convenient for comparison with experimental and other theoretical results. The relation between the light-cone helicity amplitudes, $ \mathcal{M}_{\mu'\nu',\mu\nu} $, and the c.m.s. helicity amplitudes, $\Phi_{\mu'\nu',\mu\nu}$, is as follows~\cite{Huang:2001ej},\\
\begin{align}
{\label{Eq:11}}
\Phi_{\mu'\nu',\mu\nu}=& \mathcal{M}_{\mu'\nu',\mu\nu}+\frac{\beta}{2}\left[(-1)^{\frac{1}{2}-\nu'}\mathcal{M}_{\mu'-\nu',\mu\nu}\right.\nonumber\\
                       &\left. +(-1)^{\frac{1}{2}+\nu}\mathcal{M}_{\mu'\nu',\mu-\nu}\right],
\end{align}  
where,
\begin{align}
{\label{Eq:12}}
\beta = \frac{2m}{\sqrt{s}}\frac{\sqrt{-t}}{\sqrt{s}+\sqrt{-u}}.
\end{align}  

In the handbag approximation, the WACS amplitudes, $\mathcal{M}_{\mu'\nu',\mu\nu}$, can be factored into two parts: the (hard) parton level subprocess amplitudes, $\mathcal{H}_{\mu'\nu',\mu\nu}$, and the (soft) form factors of the proton, $ R_i $, as follows~\cite{Huang:2001ej},
\begin{align}
{\label{Eq:13}}
 \mathcal{M}_{\mu'+,\mu+}(s,t) =&  \nonumber\\
& 2 \pi \alpha_{\mathrm{em}}\left[\mathcal{H}_{\mu'+,\mu+}(\hat{s},\hat{t})\left(R_V(t)+R_A(t)\right)\right. \nonumber\\
& \left. +\mathcal{H}_{\mu'+,\mu+}(\hat{s},\hat{t})\right(R_V(t)-R_A(t)\left)\right],\nonumber\\
 \mathcal{M}_{\mu' -,\mu -}(s,t) =& \nonumber\\
& \pi \alpha_{\mathrm{em}} \frac{\sqrt{-t}}{m}R_T(t)\left[\mathcal{H}_{\mu'+,\mu+}(\hat{s},\hat{t})\right. \nonumber\\
& \left. +\mathcal{H}_{\mu'+,\mu+}(\hat{s},\hat{t})\right].
\end{align}  
Note that the $\mathcal{M}$ and $\mathcal{H}$ also satisfy Eq.~(\ref{Eq:9}). We have denoted the Mandelstam variables for the photon-parton
subprocess by $ \hat{s} $, $ \hat{t} $, and $ \hat{u} $, and for the overall photon-proton reaction by $ s $, $ t $, $ u $. In the above relations, $\mu$ and $\mu'$ refer to the polarization of photons before and after interaction. The explicit helicities in the light-cone helicity amplitudes, $\mathcal{M}$, and the subprocess amplitudes, $\mathcal{H}$, represent the polarizations of the proton and active quarks, respectively. 
The hard scattering amplitudes, $\mathcal{H}_{\mu'\nu',\mu\nu}$, associated with $\gamma q\rightarrow \gamma q$ subprocess can be calculated in the perturbative QCD. To the leading order (LO) they read~\cite{Huang:2001ej},
\begin{align}
{\label{Eq:14}}
\mathcal{H}_{++,++}^{\mathrm{LO}} = 2\sqrt{\frac{\hat{s}}{-\hat{u}}},\quad\mathcal{H}_{-+,-+}^{\mathrm{LO}} = 2\sqrt{\frac{-\hat{u}}{\hat{s}}},\quad\mathcal{H}_{-+,++}^{\mathrm{LO}} = 0,
\end{align}  
while they have more complicated forms at the next-to-leading order (NLO),
\begin{align}
{\label{Eq:15}}
\mathcal{H}_{++,++}^\mathrm{NLO} &= \frac{\alpha_\mathrm{s}}{2 \pi} C_F \left\{\frac{\pi^2}{3}-7+\frac{2 \hat{t}-\hat{s}}{\hat{s}}\log\frac{\hat{t}}{\hat{u}}+\log^2\frac{-\hat{t}}{\hat{s}}\right.\nonumber\\
&\left.+\frac{\hat{t}^2}{\hat{s}^2}\left(\log^2\frac{\hat{t}}{\hat{u}}+\pi^2\right)-2i\pi \log\frac{-\hat{t}}{\hat{s}}\right\}\sqrt{\frac{\hat{s}}{-\hat{u}}},\nonumber\\
\mathcal{H}_{-+,-+}^\mathrm{NLO} &=\frac{\alpha_\mathrm{s}}{2 \pi} C_F \left\{\frac{4}{3}\pi^2-7+\frac{2\hat{t}-\hat{u}}{\hat{u}}\log\frac{-\hat{t}}{\hat{s}}+\frac{\hat{t}^2}{\hat{u}^2}\log^2\frac{-\hat{t}}{\hat{s}}\right.\nonumber\\
&\left.+\log^2\frac{\hat{t}}{\hat{u}}-2i\pi\left(\frac{2\hat{t}-\hat{u}}{2\hat{u}}+\frac{\hat{t}^2}{\hat{u}^2}\log\frac{-\hat{t}}{\hat{s}}\right)\right\}\sqrt{\frac{-\hat{u}}{\hat{s}}},\nonumber\\
\mathcal{H}_{-+,++}^\mathrm{NLO} &= -\frac{\alpha_\mathrm{s}}{2 \pi} C_F\left\{\sqrt{\frac{-\hat{u}}{\hat{s}}}+\sqrt{\frac{\hat{s}}{-\hat{u}}}\right\}.
\end{align}
Where $C_F=4/3$ is QCD color factor.
%
%
%

The soft form factors of the proton in Eq.~(\ref{Eq:13}) are denoted by $R_i$, where the subscript $i = V,A,T$ stands for vector, axial, and transverse, respectively. They can be expressed as follows,
\begin{align}
{\label{Eq:17}}
R_V^q(t) =& \int_{-1}^1 \frac{dx}{x} H^q(x,t),\nonumber\\
R_A^q(t) =& \int_{-1}^1 \frac{dx}{x}\ \mathrm{sign}(x) \widetilde{H}^q(x,t),\nonumber\\
R_T^q(t) =& \int_{-1}^1 \frac{dx}{x} E^q(x,t).
\end{align}  
Indeed, as mentioned before, the skewness dependence drops in $x$ integration due to the Lorentz invariance so that one can safely omit this variable and take the GPDs at $\xi=0$. Now, using the above relations, one can obtain the full form factors from the individual quark contributions as follows,
\begin{align}
{\label{Eq:18}}
R_i(t) = \sum_q e^2_q R_i^q(t),
\end{align}  
where $e_q$ is the charge of quark $q$ in units of positron charge. More explicitly, we have
\begin{align}
{\label{Eq:19}}
R_V(t) =& \sum_q e^2_q \int_0^1 \frac{dx}{x} [H_v^q(x,t)+2H^{\bar q}(x,t)],\nonumber\\
R_A(t) =& \sum_q e^2_q \int_0^1 \frac{dx}{x} [\widetilde{H}_v^q(x,t)+2\widetilde{H}^{\bar q}(x,t)],\nonumber\\
R_T(t) =& \sum_q e^2_q \int_0^1 \frac{dx}{x} [E_v^q(x,t)+2E^{\bar q}(x,t)].
\end{align} %
In general, the Mandelstam variables at the partonic level are different from those of the whole process. Following the work of Diehl~\textit{et~al.}~\cite{Diehl:2002ee}, and assuming massless quarks, we introduce three scenarios to relate these two sets of variables. If the mass of the proton can be neglected, the matching of the subprocess and full Mandelstam variables is simplest. In this case we have
\begin{align}
{\label{Eq:20}}
\mathrm{scenario\ 1:}\quad \hat{s}=s,\quad \hat{t}=t,\quad \hat{u}=u.
\end{align}  
In order to estimate the influence of the proton mass, two
more scenarios are also introduced:
\begin{align}  
{\label{Eq:21}}
&\mathrm{scenario\ 2:}\quad \hat{s}=s-m^2,\quad \hat{t}=t,\quad \hat{u}=u-m^2.\nonumber\\
&\mathrm{scenario\ 3:}\quad \hat{s}=s-m^2,\quad \hat{t}=-\frac{\hat{s}}{2}\left(1-\cos \theta_\mathrm{cm}\right),\nonumber\\
&\qquad\qquad\qquad \hat{u}=-\hat{s}-\hat{t}.
\end{align}  
It is worth noting that, in contrast to scenario 1, the relation $\hat s+\hat t+\hat  u=0$ holds in scenarios 2 and 3 even though we do not ignore the nucleon mass, since $s+t+u=2m_p^2 $. Note also that the aforementioned differences in the Mandelstam variables can be regarded as a source of theoretical uncertainty to the results~\cite{Diehl:2002ee}. We will study in detail the effects of considering these different scenarios for calculating the WACS cross section on the quality of the fit in Sec.~\ref{sec:four:B}.

\subsection{Modeling the GPDs}\label{sec:two:C}
As a result of many theoretical and phenomenological work, there are numerous models~\cite{Pasquini:2005dk,Pasquini:2006dv,Dahiya:2007mt,Frederico:2009fk,Mukherjee:2013yf,Maji:2015vsa} and parameterizations~\cite{Goldstein:2010gu,Goldstein:2013gra,Sharma:2016cnf} for GPDs such as Radyushkin's double distributions~\cite{Radyushkin:1997ki}, light-front constituent quark models~\cite{Frederico:2009fk}, and quark-diquark induced parameterization~\cite{Goldstein:2010gu}. Here, as in our previous work~\cite{Hashamipour:2019pgy}, we consider a simple but advantageous Ansatz suggested by DFJK~\cite{Diehl:2004cx,Diehl:2007uc,Diehl:2013xca}. This Ansatz expresses the (polarized) GPDs as a product of ordinary (PPDFs) PDFs and an exponential which contains the $ t $ dependence of GPDs, regulated with a specific profile function in $ x $. This structure is such that, in the forward limit, (polarized) GPDs will reduce to their ordinary (PPDF) PDF equivalents. For example, for positive $ x $, the GPD $ H $ changes to the usual quark and antiquark densities as $ H^q(x,0,0)=q(x) $ and $ H^q(-x,0,0)=\bar{q}(x) $. According to the DFJK Ansatz which gives $x$ and $t$ dependence of GPDs at zero skewness, the valence GPDs $ H_v^q $ can be related to ordinary valence PDFs as
\begin{equation}
H_v^q(x,t,\mu^2)= q_v(x,\mu^2)\exp [tf_q(x)],
\label{Eq:22}
\end{equation}
where $ \mu $ is the factorization scale at which the partons are resolved, just like the usual quark densities~\cite{Diehl:2004cx}. It should be noted that in Ansatz~(\ref{Eq:22}), which has a motivation from Regge theory, the profile function $ f_q(x) $ in the exponential is $ x $-dependent slope of $ \partial\left(\log H_v^q\right)/{\partial t} $ and can have various functional forms. In the simplest form, which we shall refer to as the simple Ansatz, $ f_q(x) $ is as follows,
\begin{equation}
f_q(x)=\alpha^{\prime}(1-x)\log \frac{1}{x},
\label{Eq:23}
\end{equation}
where $\alpha^{\prime}$ is an adjustable parameter that should be determined from the analysis of the relevant experimental data. In fact, it has been indicated~\cite{Diehl:2004cx} that the low- and high-$ x $ behavior of $ f_q(x) $, as well as the intermediate-$ x $ region, can be well characterized by the following forms
\begin{equation}
\label{Eq:24}
f_q(x)=\alpha^{\prime}(1-x)^2\log\frac{1}{x}+B_q(1-x)^2 + A_qx(1-x),
\end{equation}
and
\begin{equation}
\label{Eq:25}
f_q(x)=\alpha^{\prime}(1-x)^3\log\frac{1}{x}+B_q(1-x)^3 + A_qx(1-x)^2,
\end{equation}
where $ A_q $ and $ B_q $ are additional adjustable parameters. For example, the above profile functions were used for a fit to the experimental data of the Dirac and Pauli FFs~\cite{Diehl:2004cx}, and also for a phenomenological study of the strange Dirac form factor $ F_1^{s} $~\cite{Diehl:2007uc}. It is worth noting that a profile function of the form $\alpha' \log (1/x) +B$ is also used for studying the hard exclusive pion electroproduction, where GPDs play a key role~\cite{Goloskokov:2011rd}. However, in our previous work~\cite{Hashamipour:2019pgy}, we studied the effect of using this profile function for analyzing the nucleon AFF data and observed that it will not lead to an improvement in the quality of the fit. Moreover, we indicated that among the profile functions~(\ref{Eq:23}),~(\ref{Eq:24}), and~(\ref{Eq:25}), the last one can lead to lower values for the $ \chi^2 $ divided by the number of degrees of freedom, $ \chi^2 /\mathrm{d.o.f.} $, by considering the same $ A_v $ and $ B_v $ parameters for valence quarks and setting the corresponding sea quark parameters equal to zero.

According to DFJK model, an Ansatz similar to that shown in Eq.~(\ref{Eq:22}) can also be considered for the valence polarized GPDs $\widetilde{H}^q_v$, so that they can be related to valence PPDFs, $\Delta q_v(x)\equiv q^+(x)-q^-(x)$, as following
\begin{equation}
\widetilde{H}^q_v(x,t,\mu^2)=\Delta q_v(x,\mu^2) \exp [t \widetilde{f}_q(x)],
\label{Eq:26}
\end{equation}
where $\widetilde{f}_q(x)$ is the corresponding profile function which again can have a simple form like Eq.~(\ref{Eq:23}), or a complex form with more adjustable parameters like Eq.~(\ref{Eq:24}) or~(\ref{Eq:25}).

In the present work, we are going to analyze both the AFF and WACS data simultaneously to determine GPDs and also compare the results with the previous ones obtained by analyzing the AFF data solely~\cite{Hashamipour:2019pgy}, in order to investigate the impact of WACS data on the extracted densities, especially at larger values of $ -t $. To this end, we use the DFJK Ansatzes of Eqs.~(\ref{Eq:22}) and~(\ref{Eq:26}) for the unpolarized and polarized GPDs $ H $ and $ \widetilde{H} $, respectively. Moreover, for simplicity, we use these Ansatzes both for the valence and sea quark contributions as before~\cite{Hashamipour:2019pgy}, though the physical motivation to use them for the sea contributions is not as strong as that of the valence ones~\cite{Diehl:2013xca}. As is expected and we shall show, the contributions of the sea are not significant compared to those that come from the valence sector. Another point which should be mentioned is that, according to Eq.~(\ref{Eq:6}), only the polarized GPDs $ \widetilde{H} $ contribute to the AFF, while for the case of WACS cross section three kinds of GPDs, namely $ H $, $ \widetilde{H} $, and $ E $, are involved through the soft FFs of the proton $ R_V $, $ R_A $, and $ R_T $, respectively (see Eq.~(\ref{Eq:19})). Since the contribution of the $ R_T $ in WACS cross section is considerably smaller than $ R_V $ and $ R_A $ (see Fig.\ 24 of Ref.~\cite{Diehl:2013xca}), we fix the GPDs $ E_v^q(x,t) $ and $ E^{\bar q}(x,t) $ from the analysis of Ref.~\cite{Diehl:2004cx}, and just parameterize the $ H $ and $ \widetilde{H} $ GPDs. Consequently, the constraints on $ H $ come just from the WACS data, while $ \widetilde{H} $ can be constrained by both the AFF and WACS data.

\subsection{Impact parameter dependent PDFs and nucleon helicity flip distribution $ E $}\label{sec:two:D} 
As is well known, the Fourier transform of GPDs at zero skewness, $ \xi=0 $, which are called impact parameter dependent PDFs, satisfy positivity constraints so that one can associate the physical interpretation of a probability density with them. Indeed, they describe
the distribution of partons in the transverse plane~\cite{Diehl:2004cx,Burkardt:2000za,Burkardt:2002hr}. Therefore, one can relate GPDs $ H $ and $ \widetilde{H} $ to the impact parameter distribution of unpolarized quarks in an unpolarized nucleon and the distribution of longitudinally polarized quarks in a longitudinally polarized nucleon, respectively. Moreover, $ E $ is related to the distortion of the unpolarized quark distribution in the transverse
plane when the nucleon has transverse polarization. In fact, the distribution $ E_v^q(x,\xi,t) $ at zero skewness describes proton-helicity flip in a frame where the proton moves fast~\cite{Burkardt:2002hr}, i.e., in the infinite momentum frame. 

As mentioned, one can achieve a density interpretation of GPDs at $\xi=0$ in the mixed representation of
longitudinal momentum and transverse position in the infinite-momentum
frame~\cite{Burkardt:2000za,Burkardt:2002hr}. For instance, the impact parameter dependent parton distribution related to $ H $ can be defined as follows~\cite{Diehl:2004cx}
\begin{equation}        
\label{Eq:27}
q_v(x,\bm{b})= \int \frac{\mathrm{d}^2 \bm{\mathbold{\Delta}}}{(2\pi)^2}\,
\mathrm{e}^{-\mathrm{i} \bm{b}. \mathbold{\Delta} }\, H_v^q(x,t=-\mathbold{\Delta}^2).
\end{equation}
From the physical point of view, $ q_v(x,\bm{b}) $ gives the probability of finding a quark with longitudinal momentum fraction $x$ and impact parameter $\bm{b}=(b^x,b^y)$ minus the probability of finding an antiquark with the same $x$ and $\bm{b}$. Here we have adhered to the usual convention of using the $ x $ and $ y $ superscripts to denote the $ x $ and $ y $ axes in the transverse plane, where the distinction between the former and the Bjorken $ x $ is implicit. Actually, the impact parameter $\bm{b}$ in $ q_v(x,\bm{b}) $
is the transverse distance between the struck parton and
the center of momentum of the hadron~\cite{Burkardt:2002hr}. We use the boldface notation for the two-dimensional vectors in the transverse plane. We can obtain the average of impact parameter squared from $ q_v(x,\bm{b}) $ as follows~\cite{Diehl:2004cx}
\begin{equation}
\label{Eq:28}
\langle \bm{b}^2 \rangle^q_x =  \frac{\int \mathrm{d}^2\bm{b}\, \bm{b}^2\, q_v(x,\bm{b})}{
               \int \mathrm{d}^2\bm{b}\, q_v(x,\bm{b})}
      = 4 \frac{\partial}{\partial t} \log H_v^q(x,t) \Bigg|_{t=0},
\end{equation}
which is to be calculated at fixed $x$. It is worth noting that the use of an exponential Ansatz for the $ t $-dependence of unpolarized GPDs, given in Eq.~(\ref{Eq:22}), guarantees that $ q_v(x,\bm{b}) $ is positive. An estimate of the size of the hadron can be provided using the relative distance between the struck parton and the spectator system, $\bm{b}/(1-x)$. As explained in Ref.~\cite{Diehl:2004cx}, the average square of this distance is calculated from
\begin{equation}
\label{Eq:29}
d_q^2(x) = \frac{\langle \bm{b}^2 \rangle^q_x}{(1-x)^2},
\end{equation}
so that $ d_q $ provides a lower limit for the transverse size of the hadron.

It can be shown that~\cite{Burkardt:2002hr} if one changes basis from longitudinal to
transverse polarization states of the proton, $E^q_v$ has also a probability interpretation in impact parameter space. In analogy to Eq.~(\ref{Eq:27}), the probability to find an unpolarized quark with
momentum fraction $ x $ and impact parameter $\bm{b}$ in a transversly polarized proton, minus the probability to find an antiquark with the same $x$ and $\bm{b}$ is given by the distribution~\cite{Burkardt:2002hr},
\begin{equation}
\label{Eq:30}
q_{v}^{X}(x,\bm{b}) = q_v(x,\bm{b}) - \frac{b^{y}}{m}\,
\frac{\partial}{\partial \bm{b}^2}\, e_v^q(x,\bm{b}),
\end{equation}
where the Fourier transform $ e_v^q(x,\bm{b}) $ is given by,
\begin{equation}        
\label{Eq:31}
e_v^q(x,\bm{b}) = 
  \int \frac{\mathrm{d}^2 \bm{\mathbold{\Delta}}}{(2\pi)^2}\,
\mathrm{e}^{-\mathrm{i} \bm{b}.\mathbold{\Delta} }\, E_v^q(x,t=-\mathbold{\Delta}^2).
\end{equation}
Eq.~(\ref{Eq:30}) clearly indicates that transverse target polarization induces a shift in the quark distribution along the $y$-axis. One can consider the classical picture of the polarized proton as a sphere that rotates about the $x$-axis and moves in the $ z $-direction to better understand such effect~\cite{Burkardt:2002hr,Belitsky:2003nz}. The average of this shift is given by
\begin{equation}
\label{Eq:32}
\langle b^{y} \rangle^{q}_{x} 
 = \frac{\int \mathrm{d}^2\mathbold{b}\; b^{y}\, q_v^X(x,\mathbold{b})}{
               \int \mathrm{d}^2\mathbold{b}\, q_v^X(x,\mathbold{b})}
 = \frac{1}{2m}\, \frac{E_v^q(x,0)}{H_v^q(x,0)} \, .
\end{equation}
Note that, in this case, the corresponding shift for the distance between the struck quark and the
spectator system is as follows
\begin{equation}
\label{Eq:33}
s_q(x) = \frac{\langle b^{y} \rangle^{q}_{x}}{1-x} ,
\end{equation}
which is comparable to the distance function $ d_q(x) $ in Eq.~(\ref{Eq:29}). An important property of the impact parameter space distributions is that they satisfy certain inequalities~\cite{Burkardt:2003ck}, ensuring that the quark densities for various combinations of proton and quark spins are positive. By calculating the impact parameter dependent PDFs according to the above formulations, we can present the proton tomography that illustrates the interplay between longitudinal and transverse partonic degrees of freedom in the proton.

In the following, by performing various $ \chi^2 $ analyses of the AFF and WACS data at NLO, we determine the optimum values of not only the free parameters of the profile function~(\ref{Eq:25}), in particular $ \alpha^{\prime} $, but also the scale $ \mu $ at which the PDFs and PPDFs are chosen in Ansatzes~(\ref{Eq:22}) and~(\ref{Eq:26}), respectively.

It should be noted that the Regge phenomenology and various studies have indicated that the value of $ \alpha^{\prime} $ should be close to 1~\cite{Diehl:2004cx,Diehl:2007uc,Diehl:2013xca}. Although our final analysis of the AFF data~\cite{Hashamipour:2019pgy} also confirmed this value, it led to a smaller value for $ \mu $ (1 GeV) than that which has been considered in Refs.~\cite{Diehl:2004cx,Diehl:2007uc,Diehl:2013xca} (2 GeV).
So, it is of interest to investigate to what extent the inclusion of WACS data in the analysis affects the values of $ \alpha^{\prime} $ and $ \mu $, along with the shape of extracted GPDs.

%
\section{Data selection}\label{sec:three}

As mentioned before, our goal is analyzing the AFF and WACS experimental data simultaneously, in order to determine quark GPDs $ H$ and $ \widetilde{H} $ within the theoretical framework described in Sec.~\ref{sec:two}. To this aim, an important step is gathering the available experimental data for both processes. Contrary to the electromagnetic nucleon FFs which their extraction has a long history and remains a popular field of experimental research (see Ref.~\cite{Diehl:2013xca} for an overview), there are fewer measurements of the nucleon AFF. In fact, at the present, one can only use the (anti)neutrino scattering off nucleons and charged pion electroproduction to determine nucleon AFF. Most of the available measurements for the nucleon AFF obtained from charged pion electroproduction have been reviewed and discussed in Refs.~\cite{Bernard:2001rs,Schindler:2006jq}, and for the case of (anti)neutrino scattering experiments, one can refer to the data obtained from MiniBooNE experiments~\cite{Butkevich:2013vva}. Similarly, there are not many measurements of WACS cross section. As a comprehensive effort, one can refer to the measurements by the Jefferson Lab Hall A Collaboration~\cite{Danagoulian:2007gs} which include 25 kinematic settings over the range $ s=5-11 $ GeV$ ^2 $ and $ -t=2-7 $ GeV$ ^2 $.

In our previous work~\cite{Hashamipour:2019pgy}, we used the available data of nucleon AFF to determine polarized GPDs $ \widetilde{H} $ using the DFJK Ansatz of Eq.~(\ref{Eq:26}) and considering various profile functions. Although there were 84 data points from Refs.~\cite{Butkevich:2013vva,Amaldi:1970tg,Amaldi:1972vf,Bloom:1973fn,Brauel:1973cw,DelGuerra:1975uiy,DelGuerra:1976uj,Joos:1976ng,Esaulov:1978ed,Choi:1993vt,Choi:1993}, we formed a ``reduced" data set by removing data points with the same value of $ -t $ and retaining the most accurate ones which reduced the number of data to 40 (see Fig.~1 of Ref.~\cite{Hashamipour:2019pgy}). However, because of the inconsistencies in data sets (our investigations indicated that these inconsistencies are not due to the normalization), analyzing the reduced set led to a large value for the $ \chi^2 $ divided by the number of degrees of freedom, $ \chi^2 /\mathrm{d.o.f.} $, of the order of 4.5. In this work, we further reduce our AFF data by removing the older data sets and only use data from Refs.~\cite{Butkevich:2013vva,Esaulov:1978ed,DelGuerra:1975uiy,Bloom:1973fn,Joos:1976ng,Choi:1993vt}. Consequently, the number of data points of the nucleon AFF included in our new reduced set is 34 which cover the range $ 0.025 < -t < 1.84 $ GeV$ ^2 $. Another important point which should be mentioned is that we use data as $ G_A(-t) $, just like before, while the original data are given as $ G_A(-t)/G_A(0) $. In fact, as we have indicated, the quantity $ G_A(-t)/G_A(0) $ cannot put any constraint on the value of scale $ \mu $ at which PPDFs are chosen, and hence the resulting value for $ \alpha^{\prime} $ is not very reliable in such cases. For extracting  $ G_A(-t) $ from original $ G_A(-t)/G_A(0) $ data, we use the latest value of $ G_A(0) $ (axial charge $ g_A $) from PDG~\cite{Tanabashi:2018oca}, i.e. $g_A=1.2723\pm 0.0023$.

For the case of WACS data, we use the measurements by the Jefferson Lab Hall A Collaboration~\cite{Danagoulian:2007gs} with four values of $ s $, namely $ s= 4.82, 6.79, 8.90,$ and 10.92 GeV$ ^2 $. Their measurements contain 25 data points which cover a wide range of $ -t $, i.e. $ 1.65 < -t < 6.46 $ GeV$ ^2 $. As can be seen, although  the AFF data can be used to constrain
polarized GPDs mainly at smaller values of $ -t $, some important information can also be obtained from WACS data, which mainly constrain the large $ -t $ behavior of $ \widetilde{H} $. In the next section, we perform various analyses on the aforementioned AFF and WACS data simultaneously to study in detail the possible changes in the extracted GPDs, and also the values of $ \alpha^{\prime} $ and $ \mu $.

%
\section{Results and Discussions}\label{sec:four} 

In the previous sections, we have introduced the theoretical framework and experimental data which we use in the present study to determine GPDs $ H $ and $ \widetilde{H} $. In this section, we perform various analyses of the AFF and WACS data at NLO using different approaches to find the best one which leads to lowest $ \chi^2 $ with minimum number of parameters. In this regard, we first do a parameterization scan (similar to what was done in Ref.~\cite{Aaron:2009aa}) to find the optimum form of profile function~(\ref{Eq:25}) for each flavor of $ H $ and $ \widetilde{H} $ GPDs, and reduce the number of free parameters as far as possible. Then, since the theoretical calculation of WACS cross section can be done within the various scenarios (see Sec.~\ref{sec:two:B}), we perform three separate analyses in order to find the best scenario which leads to lowest $ \chi^2 $ for AFF and WACS data simultaneously, and thus the best agreement between the theoretical predictions and experimental data. As mentioned before, in our previous work~\cite{Hashamipour:2019pgy} we showed  that if $ \widetilde{H} $ is represented by the Ansatz of Eq.~(\ref{Eq:26}) for calculating $ G_A $, the final results will not be very sensitive to the choice of PPDFs set. In this work, since the theoretical calculation of WACS cross section needs $H$ and hence the unpolarized PDFs according to Ansatz~(\ref{Eq:22}), we perform various analyses using different sets of PDFs to study the sensitivity of the results to changes of PDFs. After finding the optimum form of profile function~(\ref{Eq:25}) for each flavor, the best scenario for calculating WACS cross section, and the most compatible PDFs set, we present the final results of the extracted GPDs $ H $ and $ \widetilde{H} $ along with their uncertainties. Some comparisons between the theoretical calculations obtained using the final GPDs and the experimental data included in the analysis are also presented. Finally, we present some tomography plots by calculating the impact parameter dependent PDFs introduced in Sec.~\ref{sec:two:D}.  
It should be noted that we use the CERN program \texttt{MINUIT}~\cite{James:1975dr} for doing the optimization and finding the best values of the fit parameters. Wherever needed, we use the \texttt{LHAPDF} package~\cite{Buckley:2014ana} to access a specific PDFs or PPDFs set and also the value of the strong coupling constant $ \alpha_s(\mu) $.

\subsection{Parameterization scan}\label{sec:four:A} 
As mentioned before, in our previous study~\cite{Hashamipour:2019pgy}, we used Ansatz~(\ref{Eq:26}) to analyze AFF data and determine polarized GPDs $ \widetilde{H} $. We indicated that considering a more flexible profile function like Eq.~(\ref{Eq:25}) with the same $ A_v $ and $ B_v $ parameters for valence quarks and setting the corresponding sea quark parameters equal to zero can lead to a lower value of $ \chi^2 $, as compared to the simple profile function~(\ref{Eq:23}). In this work, we are also going to include the WACS data in the analysis and determine simultaneously the unpolarized GPDs $ H $ and polarized GPDs $ \widetilde{H} $. So, it is necessary to examine the possibility of constraining a more flexible profile function using the available data and reducing the number of free parameters as far as possible. Such a parameterization scan can be done, for example, by following the procedure described in Ref.~\cite{Aaron:2009aa} which was used to find the optimum functional forms of PDFs. 

To find the optimum forms of profile function~(\ref{Eq:25}) for GPDs $ H $ and $ \widetilde{H} $, at this stage we analyze the AFF and WACS data described in Sec.~\ref{sec:three} at NLO, using scenario 1 for theoretical calculation of WACS cross section (Eq.~(\ref{Eq:20})), and also taking the \texttt{cteq6} PDFs~\cite{Pumplin:2002vw} and \texttt{NNPDFpol1.1} PPDFs~\cite{Nocera:2014gqa} in Ansatzes~(\ref{Eq:22}) and~(\ref{Eq:26}), respectively. By doing an exhaustive analytical search, we have found that the lowest $ \chi^2 /\mathrm{d.o.f.} $ occurs when one takes the following profile functions for the polarized GPDs $ \widetilde{H}_v^u $, $ \widetilde{H}_v^d $, and $ \widetilde{H}^{\bar q} $ ($ \bar{q}= \bar{u}, \bar{d} $),
\begin{align}
\label{Eq:34}
\widetilde{f}_{u_v}(x)=&\widetilde{\alpha^{\prime}}_v(1-x)^3\log\frac{1}{x}+\widetilde{B}_{u_v}(1-x)^3 + \widetilde{A}_{u_v}x(1-x)^2, \nonumber\\
\widetilde{f}_{d_v}(x)=&\widetilde{\alpha^{\prime}}_v(1-x)^3\log\frac{1}{x},\nonumber\\
\widetilde{f}_{\bar{q}}(x)=&\widetilde{\alpha^{\prime}}_v(1-x)^3\log\frac{1}{x},
\end{align}
which are obtained by considering $ \widetilde{\alpha^{\prime}}_{u_v} = \widetilde{\alpha^{\prime}}_{d_v} = \widetilde{\alpha^{\prime}}_{\bar q} =: \widetilde{\alpha^{\prime}}_{v} $, $ \widetilde{A}_{d_v} = \widetilde{B}_{d_v}=0 $, and $ \widetilde{A}_{\bar q} = \widetilde{B}_{\bar q}=0 $, and the following profile functions for the unpolarized GPDs $ H_v^u $, $ H_v^d $, and $ H^{\bar{q}} $
\begin{eqnarray}
\label{Eq:35}
f_{u_v}(x)&=&\widetilde{\alpha^{\prime}}_v(1-x)^3\log\frac{1}{x}+ \widetilde{A}_{u_v}x(1-x)^2, \nonumber\\
f_{d_v}(x)&=&\widetilde{\alpha^{\prime}}_v(1-x)^3\log\frac{1}{x}+ \widetilde{A}_{u_v}x(1-x)^2, \nonumber\\
f_{\bar{q}}(x)&=&\widetilde{\alpha^{\prime}}_v(1-x)^3\log\frac{1}{x},
\end{eqnarray}
which are obtained by considering $ \alpha^{\prime}_{u_v} = \alpha^{\prime}_{d_v} = \alpha^{\prime}_{\bar q} = \widetilde{\alpha^{\prime}}_{v} $, $ B_{u_v}=B_{d_v}=0 $, $ A_{u_v}=A_{d_v}=\widetilde{A}_{u_v} $, and $ A_{\bar q} = B_{\bar q}=0 $. With these choices, only three parameters $ \widetilde{\alpha^{\prime}}_v $, $ \widetilde{A}_{u_v} $, and $ \widetilde{B}_{u_v} $ are free and should be determined from the fit. Finally, by also considering the scale $ \mu $ as a free parameter, we have four free parameters. Note that, although increasing the number of free parameters can somewhat reduce the value of total $ \chi^2 $, it generically leads to a larger value for $ \chi^2 /\mathrm{d.o.f.}$. Hence, a major limitation at this point in time is the number of experimental data. Nevertheless, the results of the parameterization scan shown in Eq.~(\ref{Eq:34}) seem unsatisfactory in the sense that the symmetry between $\widetilde{f}_{u_v}(x)$ and $\widetilde{f}_{d_v}(x)$ is lost once the parameters $\widetilde{A}_{d_v}$ and $\widetilde{B}_{d_v}$ are set to zero. Here we adhere to the usual practice of using the results of the parameterization scan as it stands. However, at the end of Sec.~\ref{sec:four:D} we perform a new analysis by letting $ \widetilde{A}_{d_v} $ and $ \widetilde{B}_{d_v} $ be free parameters in Eq.~(\ref{Eq:34}). Upon comparing the results, we show that this has a drastic effect on $\widetilde{H}_v^d$, which we shall argue to be an improvement, with a decrease of $3\%$ in $ \chi^2$ and at a cost of only $0.7\%$ increase in $ \chi^2 /\mathrm{d.o.f.}$ We shall compare the results in detail in Sec.~\ref{sec:four:D}.

\subsection{Study of different scenarios for calculating WACS}\label{sec:four:B} 
As we explained in Sec.~\ref{sec:two:B}, there are three scenarios to relate the Mandelstam variables at partonic level to those of the whole process for calculating the WACS cross section. Using different scenarios may lead to different theoretical predictions for WACS and thus different fit results. In this section, we are going to study this issue to find the best scenario which leads to the lowest value of $ \chi^2 /\mathrm{d.o.f.} $ in analyzing the AFF and WACS data. Note that, like before, we take the \texttt{cteq6} PDFs~\cite{Pumplin:2002vw} and \texttt{NNPDFpol1.1} PPDFs~\cite{Nocera:2014gqa} in Ansatzes~(\ref{Eq:22}) and~(\ref{Eq:26}), respectively to calculate the unpolarized and polarized GPDs $ H $ and $ \widetilde{H} $.

Table~\ref{tab:chi2_com1} shows the results of three analyses of the experimental data introduced in Sec.~\ref{sec:three} that have been performed using scenarios 1, 2, and 3 described in Sec.~\ref{sec:two:B}. This table contains the list of data sets included in the analysis, along with their related observables and references. For each data set, we have presented the value of $ \chi^2 $ divided by the number of data points, $\chi^2$/$ N_{\textrm{pts.}} $. The value of $\chi^2$/d.o.f. has also been presented for each analysis in the last row of the table. As can be seen, the least $\chi^2$/d.o.f. belongs to scenario 3 which is in good agreement with scenario 2. However, scenario 1 leads to the largest $\chi^2$/d.o.f., such that its difference with the other scenarios in total $\chi^2$ is about 82.
Another point should be noted is that the scenarios 1 and 2 lead to significantly different results for AFF data from Refs.~\cite{Butkevich:2013vva} and~\cite{Esaulov:1978ed}, even though they lead to approximately the same $\chi^2$/d.o.f.. 

Table~\ref{tab:par_com1} makes a comparison between the optimum
parameters of the profile functions (\ref{Eq:34}) and (\ref{Eq:35}), in addition to the scale $ \mu^2 $ at which the PDFs and PPDFs are chosen in Ansatzes~(\ref{Eq:22}) and~(\ref{Eq:26}), respectively, obtained from three aforementioned analyses. As can be seen, there are considerable differences between the scenarios 1, 2, and 3 so that they lead to considerably different values for parameters $ \mu $ and $ \alpha^{\prime} $. Note that among these scenarios, only scenario 3 leads to a value for $ \alpha^{\prime} $ close to 1 as expected from the previous studies~\cite{Hashamipour:2019pgy,Diehl:2004cx,Diehl:2007uc,Diehl:2013xca}. However, the value of $ \mu^2 $ obtained from scenario 3 (1.5 GeV$ ^2 $) is larger than the corresponding one obtained in our previous work~\cite{Hashamipour:2019pgy} (1 GeV$ ^2 $), where we analyzed the AFF data solely to determined the polarized GPDs $ \widetilde{H} $. From now on we consider scenario 3 as the best scenario since it has led to the lowest $\chi^2$/d.o.f. (especially for the WACS data) and a more reasonable value for $ \alpha^{\prime} $ as compared to scenario 2. In the next subsection, we continue our study by investigating the sensitivity of the results to the PDFs set we choose for calculating the unpolarized GPDs using the Ansatz Eq.~(\ref{Eq:22}).

\begin{widetext}
\begin{center}
\begin{table}[th!]
\caption{The results of three analyses of the experimental data introduced in Sec.~\ref{sec:three} that have been performed using scenarios 1, 2, and 3 described in Sec.~\ref{sec:two:B}.}\label{tab:chi2_com1}
\centering
{\scriptsize 
\newcolumntype{C}[1]{>{\hsize=#1\centering\arraybackslash}X}
\centering
\begin{tabularx}{0.7\textwidth}{llc*{4}{C{2.5cm}C{2.5cm}}C{2.5cm}}

\hline \hline
Observable           & Reference                                                                   &  \multicolumn{3}{c}{ $\chi^2$/$ N_{\textrm{pts.}} $  } \\
                     
                                                             &  	   & Scenario 1                & Scenario 2               & Scenario 3   \\
 \hline
 
$G_A$                & Butkevich and Perevalov~\cite{Butkevich:2013vva}&   74.78 / 14              &   54.81 / 14             &   69.04 / 14 \\ 
                     & Del Guerra {\it et al.}~\cite{DelGuerra:1975uiy}      &   2.99 / 4                &   5.09 / 4               &   3.36 / 4   \\ 
                     & Esaulov {\it et al.}~\cite{Esaulov:1978ed}            &   31.14 / 4               &   40.75 / 4              &   33.22 / 4  \\ 
                     & Bloom {\it et al.}~\cite{Bloom:1973fn}                &   6.84 / 6                &   6.09 / 6               &   6.23 / 6   \\ 
                     & Joos {\it et al.}~\cite{Joos:1976ng}                    &   16.42 / 5               &   21.49 / 5              &   18.90 / 5  \\
                     & Choi {\it et al.}~\cite{Choi:1993vt}                   &   $0.02$ / 1              &   $0.01$ / 1             &   $0.01$ / 1 \\ 
$\frac{d\sigma}{dt}(\mathrm{WACS})$ & Danagoulian {\it et al.}~\cite{Danagoulian:2007gs}&   193.69 / 25             &   116.13 / 25            &   113.00 / 25\\
  \hline
  \hline
Total  $ \chi^2 /\mathrm{d.o.f.} $                &                    & 325.88 / 55               & 244.37 / 55               & 243.76 / 55               \\ \hline \hline\\ 
\end{tabularx}
}
\end{table}
\begin{table}[th!]
\caption{The optimum parameters of the profile functions (\ref{Eq:34}) and (\ref{Eq:35}), in addition to the scale $ \mu^2 $ at which the PDFs and PPDFs are chosen, in Ansatzes~(\ref{Eq:22}) and~(\ref{Eq:26}), respectively, obtained from three analyses of the experimental data of Table~\ref{tab:chi2_com1} that have been performed using scenarios 1, 2, and 3 described in Sec.~\ref{sec:two:B}. }\label{tab:par_com1}
\centering
{\scriptsize 
\newcolumntype{C}[1]{>{\hsize=#1\centering\arraybackslash}X}
\centering
\begin{tabularx}{0.5\linewidth}{llc*{4}{C{3cm}C{3cm}}C{3cm}C{3cm}}  \hline \hline

               \hline
               & Parameter            & Scenario 1               & Scenario 2               & Scenario 3               \\ 
               \hline 
\hline
               &$\mu^2$             &$ 0.97\pm 0.05$           &$1.91\pm 0.32$            &$1.50\pm 0.22$            \\
&  $ \widetilde{\alpha^{\prime}}_v $ &   $1.08 \pm 0.03$           &$0.73 \pm 0.02$           &$0.99\pm 0.02$            \\
               & $ \widetilde{A}_{u_v} $  &$ 3.59 \pm 0.17 $         &$8.30 \pm 0.48$           &$3.94\pm 0.16$            \\
               &  $ \widetilde{B}_{u_v} $  &$-1.83\pm 0.06$           &$-2.12\pm 0.03$           &$-1.89\pm 0.03$           \\
               \hline 
               \hline
\end{tabularx}
}
\end{table}
\end{center}
\end{widetext}

\subsection{Dependence on the PDFs}\label{sec:four:C} 
In our previous work~\cite{Hashamipour:2019pgy}, we studied the impact of choosing different sets of PPDFs on the theoretical calculations of the nucleon AFF of Eq.~(\ref{Eq:6}), using the Ansatz~(\ref{Eq:26}), and showed that the final results were not very sensitive to the choice of PPDFs set. To be more precise, we concluded that the difference between the results obtained for $ G_A $ using the \texttt{DSSV08}~\cite{deFlorian:2009vb} and \texttt{NNPDFpol1.1}~\cite{Nocera:2014gqa} PPDFs was approximately 2\% in full range of $ -t $ under consideration. In this work, we have also included some new data from the WACS cross section which their theoretical calculations require PDFs according to Eq.~(\ref{Eq:19}) (through $ H $ in $ R_V $). Therefore, it is also of interest to study the sensitivity of the results to the PDFs set that we choose for calculating the GPDs $ H $, and subsequently the WACS cross section. 

In the previous subsections, we used the \texttt{cteq6} PDFs~\cite{Pumplin:2002vw} for doing the parameterization scan and finding the best scenario for calculating the WACS cross section. Now, in order to investigate the effect of PDFs on the fit results, in particular the value of the $ \chi^2 $, the shape of GPDs, and the optimum values of the free parameters, we repeat the analysis of Sec.~\ref{sec:four:B} using scenario 3, but this time considering more recent sets of PDFs. To this aim, we use the NLO PDFs from the \texttt{CT14}~\cite{Dulat:2015mca}, \texttt{MMHT14}~\cite{Harland-Lang:2014zoa} and \texttt{NNPDF3.0}~\cite{Ball:2014uwa} and compare their results with each other. The results of these three analyses have been presented in the third, fourth and fifth columns of Table~\ref{tab:chi2_com2}, respectively. Overall, we conclude that there are no significant differences between the analyses performed using different PDFs, i.e., the value of total $ \chi^2 $ does not change by more than 1 unit. Table~\ref{tab:par_com2} makes a comparison between the optimum parameters obtained from the three aforementioned analyses. As can be seen, the values of the parameters also do not change significantly by choosing different sets of PDFs. However, since the analysis performed using the \texttt{CT14} PDFs has led to the lowest $\chi^2$/d.o.f. according to Table~\ref{tab:chi2_com2}, we consider it to be the most suitable. Therefore, our final GPDs are those that have been obtained using scenario 3 for calculating the WACS cross section, and the \texttt{CT14} PDFs and \texttt{NNPDFpol1.1} PPDFs~\cite{Nocera:2014gqa} in Ansatzes~(\ref{Eq:22}) and~(\ref{Eq:26}), respectively. Note that, according to Table~\ref{tab:par_com2}, our new analysis, which includes the WACS data, confirms the result obtained in our previous work~\cite{Hashamipour:2019pgy} (where we analyzed the AFF data solely) and also other studies~\cite{Diehl:2004cx,Diehl:2007uc,Diehl:2013xca} for $ \alpha^{\prime} $ ($ \alpha^{\prime} \sim 1$). However, the final value of $ \mu^2 $, $ \mu^2=1.48 $ GeV$ ^2 $ is larger than our previous work (1 GeV$ ^2 $) and smaller than other studies (4 GeV$ ^2 $)~\cite{Diehl:2004cx,Diehl:2007uc,Diehl:2013xca}. 
\begin{widetext}
\begin{center}
\begin{table}[th!]
\caption{The results of three analyses of the experimental data introduced in Sec.~\ref{sec:three} that have been performed using scenario 3 described in Sec.~\ref{sec:two:B} and the \texttt{CT14}~\cite{Dulat:2015mca}, \texttt{MMHT14}~\cite{Harland-Lang:2014zoa}, and \texttt{NNPDF3.0}~\cite{Ball:2014uwa} PDFs.}\label{tab:chi2_com2}
\centering
{\scriptsize 
\newcolumntype{C}[1]{>{\hsize=#1\centering\arraybackslash}X}
\centering
\begin{tabularx}{0.7\textwidth}{llc*{4}{C{2.5cm}C{2.5cm}}C{2.5cm}}

\hline \hline
Observable           & Reference                                                                   &  \multicolumn{3}{c}{ $\chi^2$/$ N_{\textrm{pts.}} $  } \\
                     
                                                             &  	   & \texttt{CT14}                      & \texttt{MMHT14}                    & \texttt{NNPDF3.0}                    \\
 \hline

$G_A$                & Butkevich and Perevalov~\cite{Butkevich:2013vva}&   68.83 / 14              &   69.98 / 14              &   69.86 / 14              \\ 
                     & Del Guerra et al.~\cite{DelGuerra:1975uiy}      &   3.37 / 4                &   3.31 / 4                &   3.33 / 4                \\ 
                     & Esaulov et al.~\cite{Esaulov:1978ed}            &   33.32 / 4               &   32.81 / 4               &   32.86 / 4               \\ 
                     & Bloom et al.~\cite{Bloom:1973fn}                &   6.21 / 6                &   6.31 / 6                &   6.29 / 6                \\ 
                     & Joos et al\cite{Joos:1976ng}                    &   19.01 / 5               &   18.52 / 5               &   18.57 / 5               \\
                     & Choi et al.\cite{Choi:1993vt}                   &   $0.01$ / 1              &   $0.01$ / 1              &   $0.01$ / 1              \\ 
$\frac{d\sigma}{dt}(\mathrm{WACS})$ & Danagoulian et al.~\cite{Danagoulian:2007gs}&   111.77 / 25             &   112.29 / 25             &   112.48 / 25             \\  \hline

  \hline
Total $ \chi^2 /\mathrm{d.o.f.} $               &                    & 242.52 / 55               & 243.23 / 55               & 243.40 / 55               \\ \hline \hline\\ 
\end{tabularx}
}
\end{table}
\begin{table}[th!]
\caption{The optimum parameters of the profile functions (\ref{Eq:34}) and (\ref{Eq:35}), in addition to the scale $ \mu^2 $ at which the PDFs and PPDFs are chosen in Ansatzes~(\ref{Eq:22}) and~(\ref{Eq:26}), respectively, obtained from three analyses of the experimental data of Table~\ref{tab:chi2_com2} that have been performed using scenario 3 described in Sec.~\ref{sec:two:B} and the \texttt{CT14}~\cite{Dulat:2015mca}, \texttt{MMHT14}~\cite{Harland-Lang:2014zoa} and \texttt{NNPDF3.0}~\cite{Ball:2014uwa} PDFs. }\label{tab:par_com2}
\centering
{\scriptsize 
\newcolumntype{C}[1]{>{\hsize=#1\centering\arraybackslash}X}
\centering
\begin{tabularx}{0.5\linewidth}{llc*{4}{C{3cm}C{3cm}}C{3cm}C{3cm}}  \hline \hline

               \hline
               & Parameter            & \texttt{CT14}                      & \texttt{MMHT14}                    & \texttt{NNPDF3.0} \\ 
               \hline 
\hline
               &$\mu^2$             &$ 1.48\pm 0.21$           &$ 1.52\pm 0.24$           &$ 1.55\pm 0.24$            \\
&  $ \widetilde{\alpha^{\prime}}_v $ &   $0.99 \pm 0.02$           &$0.99 \pm 0.02$           &$0.99\pm 0.02$            \\
               & $ \widetilde{A}_{u_v} $  &$ 3.90 \pm 0.16 $         &$3.96 \pm 0.15$           &$3.89\pm 0.15$            \\
               &  $ \widetilde{B}_{u_v} $  &$-1.88\pm 0.03$           &$-1.90\pm 0.04$           &$-1.87\pm 0.04$           \\
               \hline 
               \hline
\end{tabularx}
}
\end{table}
\end{center}
\end{widetext}

\begin{figure}
\resizebox{!}{0.56\textwidth}
{
\input{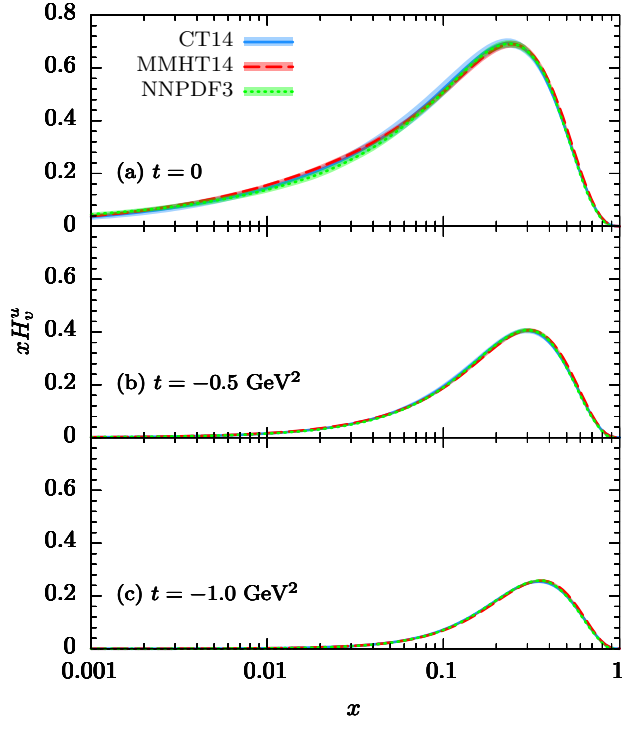}
}
\resizebox{!}{0.56\textwidth}
{
\input{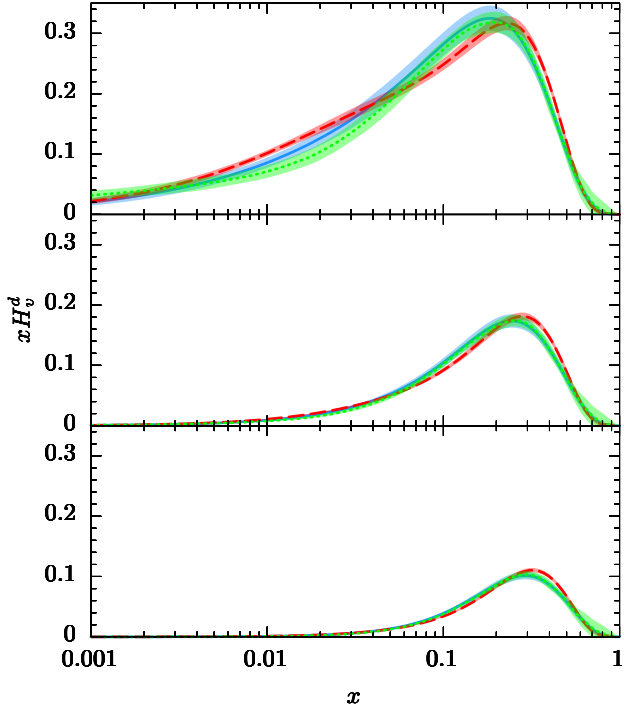}
}
\caption{A comparison between the up $ H_v^u $ (top panel) and down $ H_v^d $ (bottom panel) unpolarized GPDs with their uncertainties obtained using three different sets of NLO PDFs, namely \texttt{CT14} (blue solid curve), \texttt{MMHT14} (red dashed curve) and \texttt{NNPDF3.0} (green dotted curve), and profile function $ f_{u_v} $ of Eq.~(\ref{Eq:35}) with parameters listed in Table~\ref{tab:par_com2} at three different values of $ t $, $ t=0, -0.5 $ and $ -1 $ GeV$ ^2 $.}
\label{fig:fig2}
\end{figure}

\begin{figure}
\resizebox{!}{0.56\textwidth}
{
\input{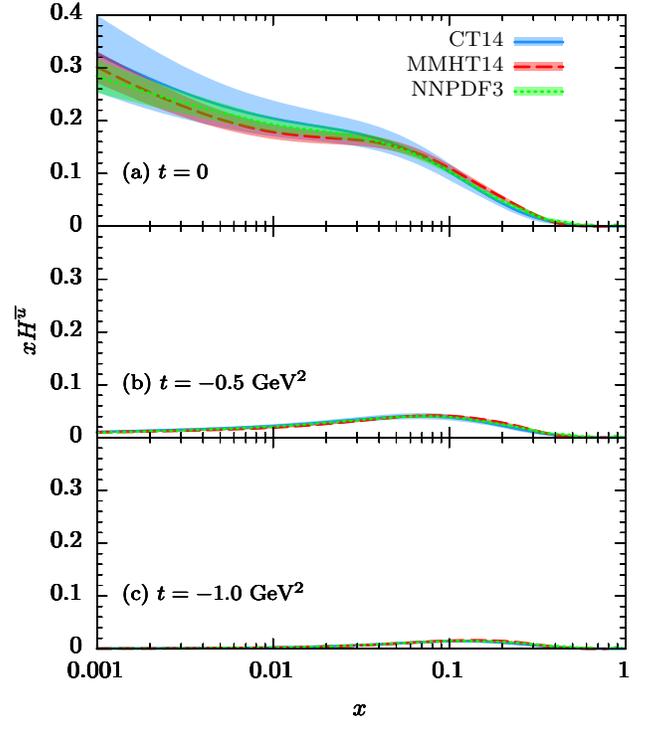}
}
\resizebox{!}{0.56\textwidth}
{
\input{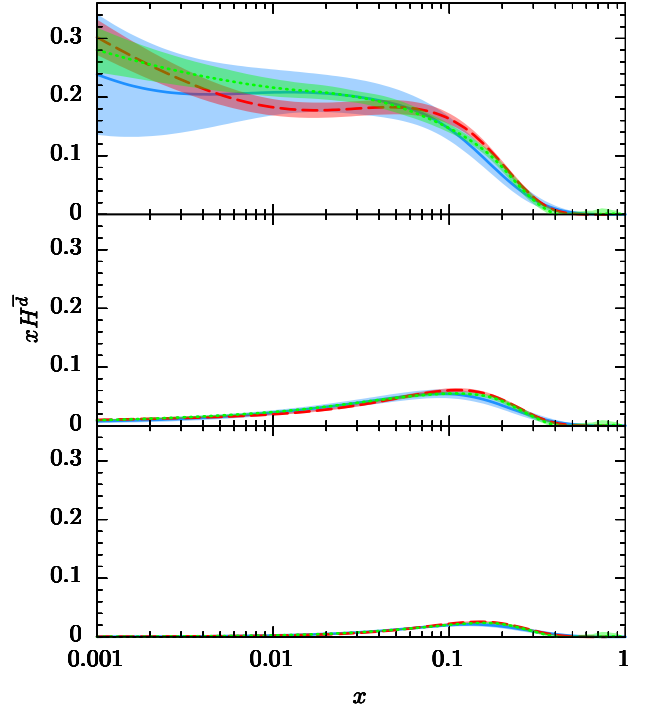}
}
\caption{Same as Fig.~\ref{fig:fig2} but for the up and down sea quark GPDs $ H^{\bar{u}} $ (top panel) and $ H^{\bar{d}} $ (bottom panel).}
\label{fig:fig3}
\end{figure}

Although there are no significant differences between the results of analyses performed using different sets of PDFs in view of $ \chi^2 $ (Table~\ref{tab:chi2_com2}) and parameter (Table~\ref{tab:par_com2}) values, it is also of interest to compare different GPDs of various flavors obtained from three aforementioned analyses. Top panel of Fig.~\ref{fig:fig2} shows a comparison between the $ H_v^u $ GPDs with their uncertainties obtained using three different sets of NLO PDFs, namely \texttt{CT14} (blue solid curve), \texttt{MMHT14} (red dashed curve) and \texttt{NNPDF3.0} (green dotted curve), and profile function $ f_{u_v} $ of Eq.~(\ref{Eq:35}) with parameters listed in Table~\ref{tab:par_com2} at three different values of $ t $, $ t=0, -0.5 $ and $ -1 $ GeV$ ^2 $. As can be seen, the differences between these three sets of GPDs are not very significant. Indeed, we have evaluated them and found that, excluding the regions close to the end points, the differences are less than 5\%. However, the results differ more for $ H_v^d $ GPDs which have been compared in the bottom panel of Fig.~\ref{fig:fig2}. In fact, in this case, the differences can reach 20\% in the same domain as above. An important point which should be mentioned is that, according to Eq.~(\ref{Eq:35}), the profile functions $ f_{u_v} $ and $ f_{d_v} $ are equal. Therefore, the larger differences observed in the bottom panel of Fig.~\ref{fig:fig2} between the $ H_v^d $ GPDs compared to the $ H_v^u $ GPDs in the top panel come directly from the larger differences in $ d_v $ PDFs compared to $ u_v $ PDFs of the \texttt{CT14}, \texttt{MMHT14} and \texttt{NNPDF3.0} sets. Figure \ref{fig:fig3} shows the same comparison of Fig.~\ref{fig:fig2} but for up and down sea quark GPDs $ H^{\bar{u}} $ (top panel) and $ H^{\bar{d}} $ (bottom panel). As can be seen, the three sets of PDFs produce larger differences for the $H^{\bar{u}}$ and $H^{\bar{d}}$, as compared to their valance counterparts.

\begin{figure}
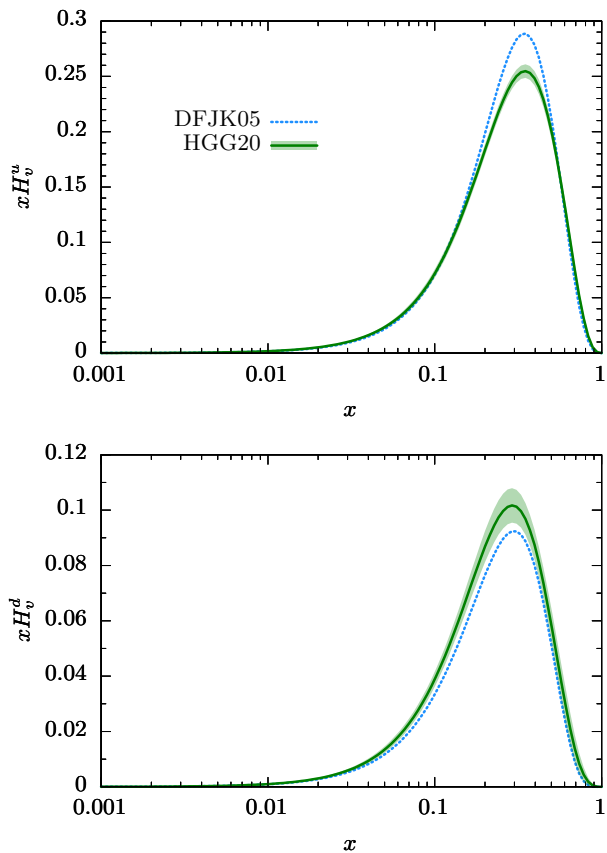

\resizebox{!}{0.32\textwidth}
{
\input{./fig-huv_DFJK.tex}
}
\resizebox{!}{0.32\textwidth}
{
\input{./fig-hdv_DFJK.tex}
}
\caption{A comparison between our final unpolarized GPDs (green solid curve labeled as HGG20) for the up (top panel) and down (bottom panel) valence quarks with the results obtained by Diehl, Feldmann, Jakob, and Kroll (blue dotted curve labeled as DFJK05)~\cite{Diehl:2004cx}. See Sec.~\ref{sec:four:C} for more information.}
\label{fig:fig4}
\end{figure}

As mentioned before, we consider the GPDs obtained using scenario 3 for calculating the WACS cross section, and the \texttt{CT14} PDFs and \texttt{NNPDFpol1.1} PPDFs~\cite{Nocera:2014gqa} in Ansatzes~(\ref{Eq:22}) and~(\ref{Eq:26}), respectively, as our final GPDs. To check the validity of the results obtained, in Fig.~\ref{fig:fig4} we have compared our GPDs (green solid curve labeled as HGG20) for the up (top panel) and down (bottom panel) valence quarks with the results obtained by Diehl, Feldmann, Jakob, and Kroll (blue dotted curve labeled as DFJK05)~\cite{Diehl:2004cx}. This figure clearly shows a good agreement between the results of these two analyses. Note that in the DFJK05 analysis, the main body of the experimental data was composed of the Dirac and Pauli form factors of the nucleon, while our analysis contains data from the AFF and WACS.
\subsection{Final polarized GPDs}\label{sec:four:D} 
\begin{figure}
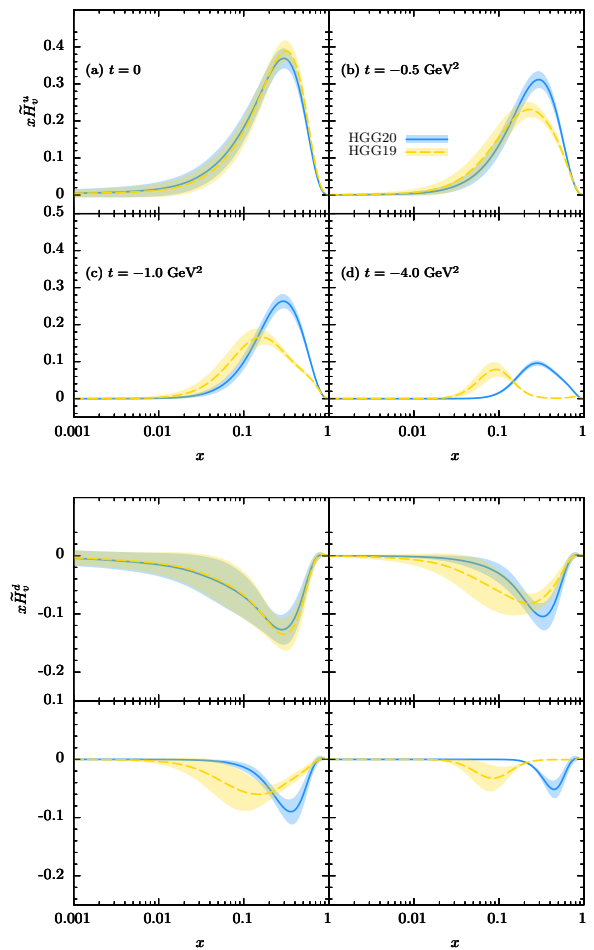

\resizebox{!}{0.36\textwidth}
{
\input{./fig-ht_uv.tex}
}
\resizebox{!}{0.36\textwidth}
{
\input{./fig-ht_dv.tex}
}
\caption{A comparison between our final results for the polarized GPDs $ \widetilde{H}_v^u $ (top panel) and $ \widetilde{H}_v^d $ (bottom panel) with their uncertainties obtained from this work (blue solid curve labeled as HGG20) and our previous work (orange dashed curve labeled as HGG19)~\cite{Hashamipour:2019pgy} at four different values of $ t $, $ t=0, -0.5, -1 $ and $ -4 $ GeV$ ^2 $.}
\label{fig:fig5}
\end{figure}

In the previous subsection, we studied in detail the sensitivity of the fit results to the PDFs set that we choose for calculating the GPDs $ H $ and WACS cross section. We showed that although there are no significant differences between the analyses performed using different sets of PDFs in view of $ \chi^2 $ and parameter values, the extracted GPDs, especially for the case of down valence and sea quark distributions, differ as functions of $x$. In this subsection, we present our final results for the polarized GPDs $ \widetilde{H}_v^u $, $ \widetilde{H}_v^d $, and $ \widetilde{H}^{\bar{q}} $ which can be calculated using the Ansatz (\ref{Eq:26}) and profile functions (\ref{Eq:35}) with parameter values listed in Table~\ref{tab:par_com2} for the \texttt{CT14} analysis and \texttt{NNPDFpol1.1} as the PPDFs set. Figure~\ref{fig:fig5} shows a comparison between the $ \widetilde{H}_v^u $ (top panel) and $ \widetilde{H}_v^d $ (bottom panel) GPDs with their uncertainties obtained from this work (blue solid curve labeled as HGG20) and our previous work (orange dashed curve labeled as HGG19)~\cite{Hashamipour:2019pgy} at four different values of $ t $, $ t=0, -0.5, -1 $ and $ -4 $ GeV$ ^2 $. As can be seen, for both $ \widetilde{H}_v^u $ and $ \widetilde{H}_v^d $ GPDs, the results of HGG20 and HGG19 analyses are very close at $ t=0 $. However, as the absolute value of $ t $ increases, the differences between these two analyses become more significant, i.e., the peaks of HGG20 GPDs moves to the larger values of $ x $ and their magnitudes increase as compared to the HGG19 GPDs. In fact, since the HGG20 analysis has included the WACS data, in addition to the AFF data, which contain data points with larger values of $ -t $ up to 6.46 GeV$ ^2 $, such changes in valence GPDs at large $ -t $ are not unexpected. Another point that should be noted is the relative reduction in uncertainties of the HGG20 GPDs compared to the HGG19 GPDs. 

Although the inclusion of WACS data in the analysis of GPDs leads to significant changes in the shapes of extracted polarized valence GPDs at larger values of $ -t $, the situation is different for the case of up and down sea quark GPDs $ \widetilde{H}^{\bar{u}} $ and $ \widetilde{H}^{\bar{d}} $, since their contributions in the theoretical calculations of AFF and WACS cross section are small compared with the valence ones~\cite{Diehl:2004cx,Diehl:2013xca}. Figure~\ref{fig:fig6} shows the same comparison as Fig.~\ref{fig:fig5} but for the $ \widetilde{H}^{\bar{u}} $ (top panel) and $ \widetilde{H}^{\bar{d}} $ (bottom panel) GPDs. As can be seen, in this case, there are no considerable differences between the HGG20 and HGG19 GPDs even at larger values of $ -t $. This indicates that both AFF and WACS data lead to similar behavior for the sea quark polarized GPDs.
\begin{figure}
\resizebox{!}{0.36\textwidth}
{
\input{./fig-ht_ubar.tex}
}
\resizebox{!}{0.36\textwidth}
{
\input{./fig-ht_dbar.tex}
}
\caption{Same as Fig.~\ref{fig:fig5} but for the up and down sea quark GPDs $ \widetilde{H}^{\bar{u}} $ (top panel) and $ \widetilde{H}^{\bar{d}} $ (bottom panel).}
\label{fig:fig6}
\end{figure}

\begin{figure}
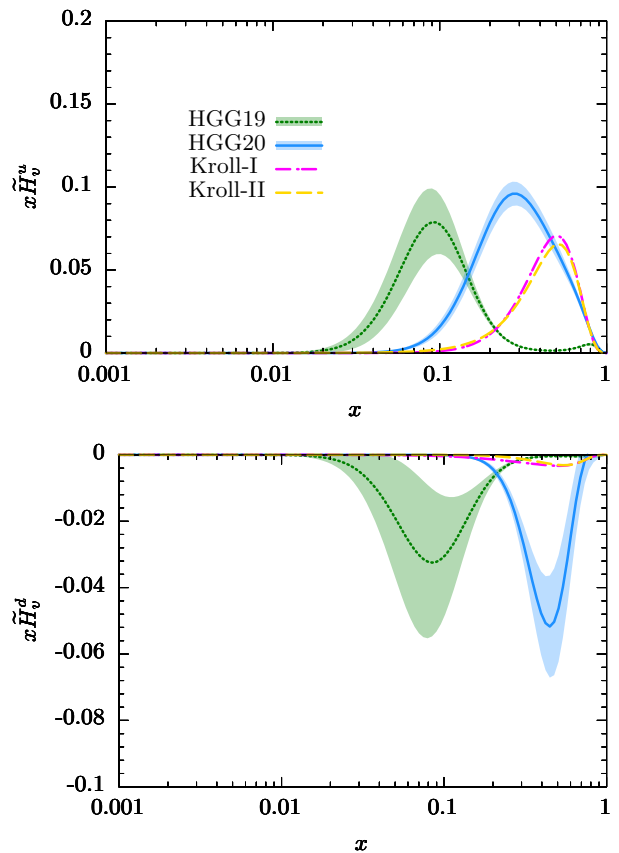

\resizebox{!}{0.32\textwidth}
{
\input{./fig-huv_triple.tex}
}
\resizebox{!}{0.32\textwidth}
{
\input{./fig-hdv_triple.tex}
}
\caption{A comparison between the results of four analyses HGG19 (dotted curve), Kroll-I (dashed curve), Kroll-II (dotted-dashed curve), and HGG20 (solid curve) for the polarized GPDs $ \widetilde{H}_v^u $ (top panel) and $ \widetilde{H}_v^d $ (bottom panel) at $ t= -4 $ GeV$ ^2 $. See Sec.~\ref{sec:four:D} for more information.}
\label{fig:fig7}
\end{figure}

In the Introduction, we compared two sets of polarized GPDs $ \widetilde{H}_v^u $ and $ \widetilde{H}_v^d $ (see Fig.~\ref{fig:fig1}) from our previous work~\cite{Hashamipour:2019pgy} (HGG19) which analyzed the AFF data solely and Kroll's work~\cite{Kroll:2017hym} which considered only the WACS data. We showed that there are considerable differences between these two sets of GPDs at large value of $ -t $ (4 GeV$ ^2 $) so that the Kroll's results are more inclined to larger $ x $. Therefore, it is of interest now to compare our new polarized GPDs (HGG20) with the HGG19 and Kroll's ones since we have extracted them using a simultaneous analysis of AFF and WACS data. Fig.~\ref{fig:fig7} shows such a comparison between the results obtained for the $ \widetilde{H}_v^u $ (top panel) and $ \widetilde{H}_v^d $ (bottom panel) GPDs from four analyses HGG19 (dotted curve), Kroll-I (dashed curve), Kroll-II (dotted-dashed curve), and HGG20 (solid curve) at $ t= -4 $ GeV$ ^2 $. This figure indicates that the simultaneous analysis of AFF and WACS data leads to a significant shift of the valence polarized GPDs to the large $ x $ region at larger values of $ -t $. Indeed, one can clearly see the better agreement of our results with those of Kroll after the inclusion of the WACS data in the analysis. Kroll did not report uncertainty for his distribution, we guess that his uncertainty should be same order as ours and as a result the uncertainty bands would touch. From the bottom panel of Fig.~\ref{fig:fig7} we infer that the inclusion of the AFF data affects significantly $ \widetilde{H}_v^d $, since both analyses containing these data (HGG19 and HGG20) have a distribution which is greater by an order of magnitude than the Kroll-I and Kroll-II.

As mentioned in Sec.~\ref{sec:four:A}, it is also of interest to study the impact of using a more flexible profile function for $\widetilde{H}_v^d$ by letting $\widetilde{A}_{d_v}$ and $\widetilde{B}_{d_v}$ be free parameters, leading to symmetrical Ansatzes for $\widetilde{H}_v^u$ and $\widetilde{H}_v^d$.
By performing a new analysis of the AFF and WACS data using scenario 3 for calculating the WACS cross section, and the \texttt{CT14} PDFs and \texttt{NNPDFpol1.1} PPDFs in Ansatzes~(\ref{Eq:22}) and~(\ref{Eq:26}), respectively, and also adding two free parameters $ \widetilde{A}_{d_v} $ and $ \widetilde{B}_{d_v} $ in profile function $ \widetilde{f}_{d_v}(x) $ of Eq.~(\ref{Eq:34}), the optimum values of the fit parameters are obtained as follows,
\begin{eqnarray*}
&\widetilde{\alpha^{\prime}}_v=0.98\pm 0.02,  \qquad \mu^2 = 2.22 \pm 0.82, & \\ 
& \widetilde{A}_{u_v}=3.79\pm 0.22,  \qquad  \widetilde{B}_{u_v}= -1.82\pm 0.05,  & \\
& \widetilde{A}_{d_v}=9.85\pm 3.43,  \qquad  \widetilde{B}_{d_v}= -2.48\pm 0.43,  &
\end{eqnarray*}
and the value of $ \chi^2 /\mathrm{d.o.f.} $ increases by 0.03 (from 4.41 to 4.44), though the value of $ \chi^2 $ itself decreases by about 7 units (from 242.5 to 235.1). Comparing these values with the corresponding ones from Table~\ref{tab:par_com2} (see column entitled \texttt{CT14}), one observes that the differences between the $\widetilde{\alpha^{\prime}}_v$, $\widetilde{A}_{u_v}$, and $\widetilde{B}_{u_v}$ parameters are not significant. However, the value of $ \mu^2 $ has increased considerably from 1.48 to 2.22 GeV$ ^2 $. By calculating $ \widetilde{H}_v^u(x) $ and  $ \widetilde{H}^{\bar q}(x) $ using the new values obtained for their parameters and comparing them with the previous distributions at typical values of $ -t $, we have found that their graphs have not changed significantly by adding two free parameters for $ \widetilde{H}_v^d(x) $. However, their uncertainties have increased, as the numbers reported above for the parameters clearly indicate. It is worth noting that the uncertainties obtained for the two new parameters $ \widetilde{A}_{d_v} $ and $ \widetilde{B}_{d_v} $ of $ \widetilde{H}_v^d(x) $ are considerably larger than those of the corresponding parameters of $ \widetilde{H}_v^u(x) $.

\begin{figure}
\resizebox{!}{0.32\textwidth}
{
\input{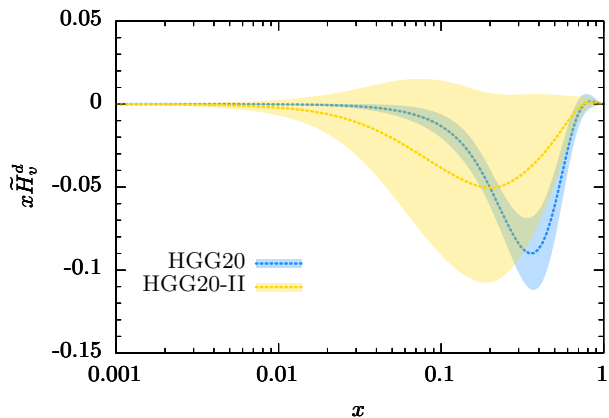}
}
\caption{A comparison between the $ \widetilde{H}_v^d(x) $ obtained from two analyses HGG20 (blue solid curve) and HGG20-II (orange dashed curve) at $ -t=1 $ GeV$ ^2 $. See text for details.}
\label{fig:fig7prime}
\end{figure}

Figure~\ref{fig:fig7prime} shows a comparison between the $ \widetilde{H}_v^d(x) $ obtained from the analysis performed by adding two free parameters $ \widetilde{A}_{d_v} $ and $ \widetilde{B}_{d_v} $ in profile function $ \widetilde{f}_{d_v}(x) $ of Eq.~(\ref{Eq:34}) labeled as ``HGG20-II" and the corresponding one from the previous analysis (HGG20) at $ -t=1 $ GeV$ ^2 $. As can be seen, there are significant differences between the results of these two analyses. To be more precise, the peak value has decreased by about a factor of two, and its position has shifted from about $x=0.3$ to $x=0.1$. If we superimpose Fig.~\ref{fig:fig7prime} onto Fig.~\ref{fig:fig7} and compare the trends of changes in the five graphs, we can see that HGG20-II, having equivalent Ansatzes for the up and down valance quarks, is more appropriate. However the increase in the uncertainty is substantial, indicating that the available data is insufficient to fully support this case.
\subsection{Data-theory comparison}\label{sec:four:E} 
\begin{figure}
\resizebox{!}{0.32\textwidth}
{
\input{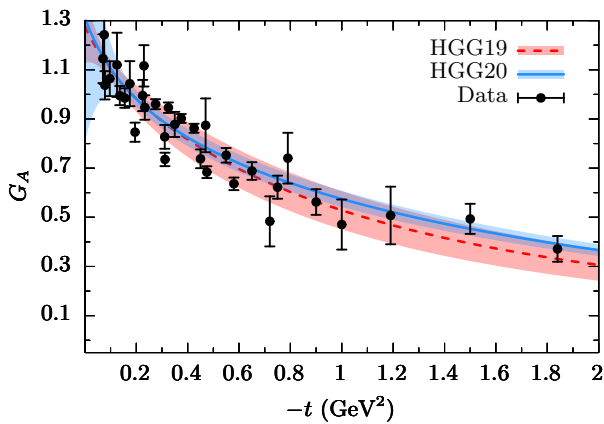}
}
\caption{A comparison between the theoretical
predictions of the $ G_A $ with their uncertainties obtained using the HGG20 (blue solid curve) and HGG19 (red dashed curve) GPDs and the experimental data listed in Table~\ref{tab:chi2_com2}.}
\label{fig:fig8}
\end{figure}

Now it is time to compare the theoretical predictions of the AFF and WACS cross section obtained using the final GPDs presented in the previous subsections with the experimental data included in the analysis in order to check the validity of the fit. Figure~\ref{fig:fig8} shows a comparison between the theoretical
predictions for the nucleon axial form factor $ G_A $ with their uncertainties obtained using the HGG20 (blue solid curve) and HGG19 red dashed curve) GPDs and the experimental data included in the analysis (see Sec.~\ref{sec:three} and Table~\ref{tab:chi2_com2}). Note again that the HGG20 prediction has been calculated using the polarized GPDs $ \widetilde{H} $ obtained from the analysis considering the \texttt{CT14} as PDFs set and \texttt{NNPDFpol1.1} as PPDFs set. As can be seen, the agreement between the theory and data has improved in the HGG20 analysis compared to our previous work (HGG19), where we analyzed the AFF data solely. Another point that should be noted is a considerable reduction in the error band of the HGG20 prediction compared to the HGG19, for $ x \gtrsim 0.2 $, though it is wider for smaller values of $ x $.

\begin{figure}
\resizebox{!}{0.32\textwidth}
{
\input{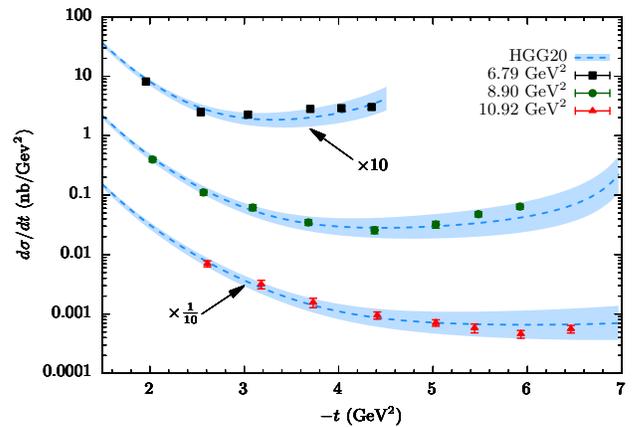}
}
\caption{A comparison between the experimental data of the WACS cross section ($ d\sigma/dt $) with the related theoretical predictions obtained using our final GPDs. The data points belong to three different values of $ s $, namely $ s=6.79, 8.90, $ and 10.92 GeV$ ^2 $, which have been shown by the square, circle, and triangle symbols, respectively. Multiplication factors indicated are to distinguish the graphs.}
\label{fig:fig9}
\end{figure}

In Fig.~\ref{fig:fig9} we have presented a comparison between the experimental data of the WACS cross section ($ d\sigma/dt $) included in the analysis with the related theoretical predictions obtained using our final unpolarized ($ H $) and polarized ($ \widetilde{H} $) GPDs and considering scenario 3 of Eq.~(\ref{Eq:21}). The data points belong to three different values of $ s $, namely $ s=6.79, 8.90, $ and 10.92 GeV$ ^2 $, which have been shown by the square, circle, and triangle symbols, respectively. The vertical error bars contain both the statistical and systematic uncertainties added in quadrature. It should be noted that in order to better distinguish between the results of these three values of $ s $, we have multiplied the experimental data and theoretical predictions of the first and third by a factor of 10 and 1/10, respectively. This figure clearly shows a good agreement between the theoretical predictions and experimental data in the entire interval of $ -t $ under consideration, and thus the good quality of the fit. Indeed, almost all data points are within the error band of the theoretical predictions. According to this figure, the reason for the relatively large $\chi^2$/$ N_{\textrm{pts.}} $ of these data (112/25 for the \texttt{CT14} analysis in Table~\ref{tab:chi2_com2}) is not the poor theoretical description of the data, but it is due to the small values of the experimental uncertainties which appear in the denominator of the expression for $ \chi^2 $.
%

\subsection{Proton tomography}\label{sec:four:F} 
As we explained in Sec.~\ref{sec:two:D}, one can associate the physical interpretation of a probability density with the impact parameter dependent PDFs, which are the Fourier transforms of GPDs at zero skewness. In this subsection, we calculate the impact parameter dependent PDFs and present proton tomography in impact parameter space using our final GPDs obtained from the simultaneous analysis of the AFF and WACS data.

\begin{figure}
\resizebox{!}{0.32\textwidth}
{
\input{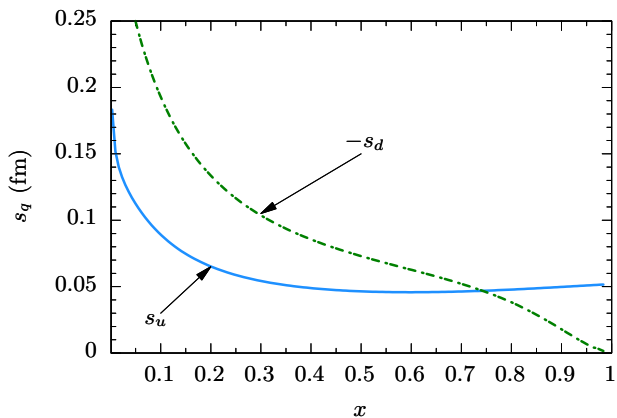}
}
\caption{Average shift $ s_q(x) $ of the distance between struck quark and
spectators in a polarized proton, defined in Eq.~(\ref{Eq:33}), for the up (blue solid curve) and down (green dotted-dashed curve) quarks.}
\label{fig:fig10}
\end{figure}

In Fig.~\ref{fig:fig10}, we have plotted the average shift $ s_q(x) $ of the
distance between struck quark and spectators in a transversely polarized proton for the up (blue solid curve) and down (green dotted-dashed curve) quarks, obtained using Eq.~(\ref{Eq:33}). Note that for the case of down quark, since the values of $ s $ are negative, we have presented the results as $ -s_d(x) $ to make the comparison more clear. In fact, the different signs of $ s_u(x) $ and $ s_d(x) $ come from the different signs of the anomalous magnetic moments of the up and down quarks~\cite{Diehl:2004cx}.
Note that, according to Eqs.~(\ref{Eq:32}) and~(\ref{Eq:33}), one needs both $ H_v^q $ and $ E_v^q $ GPDs for calculating $ s_q(x) $. As mentioned before, in our analysis, we have fixed $ E_v^q $ from the analysis of Ref.~\cite{Diehl:2004cx} and just parameterized the $ H $ and $ \widetilde{H} $ GPDs. Therefore, for calculating $ s_q(x) $ in Fig.~\ref{fig:fig10}, we have used our final result for $ H_v^q $ and the result of Ref.~\cite{Diehl:2004cx} for $ E_v^q $. 

\begin{figure*}
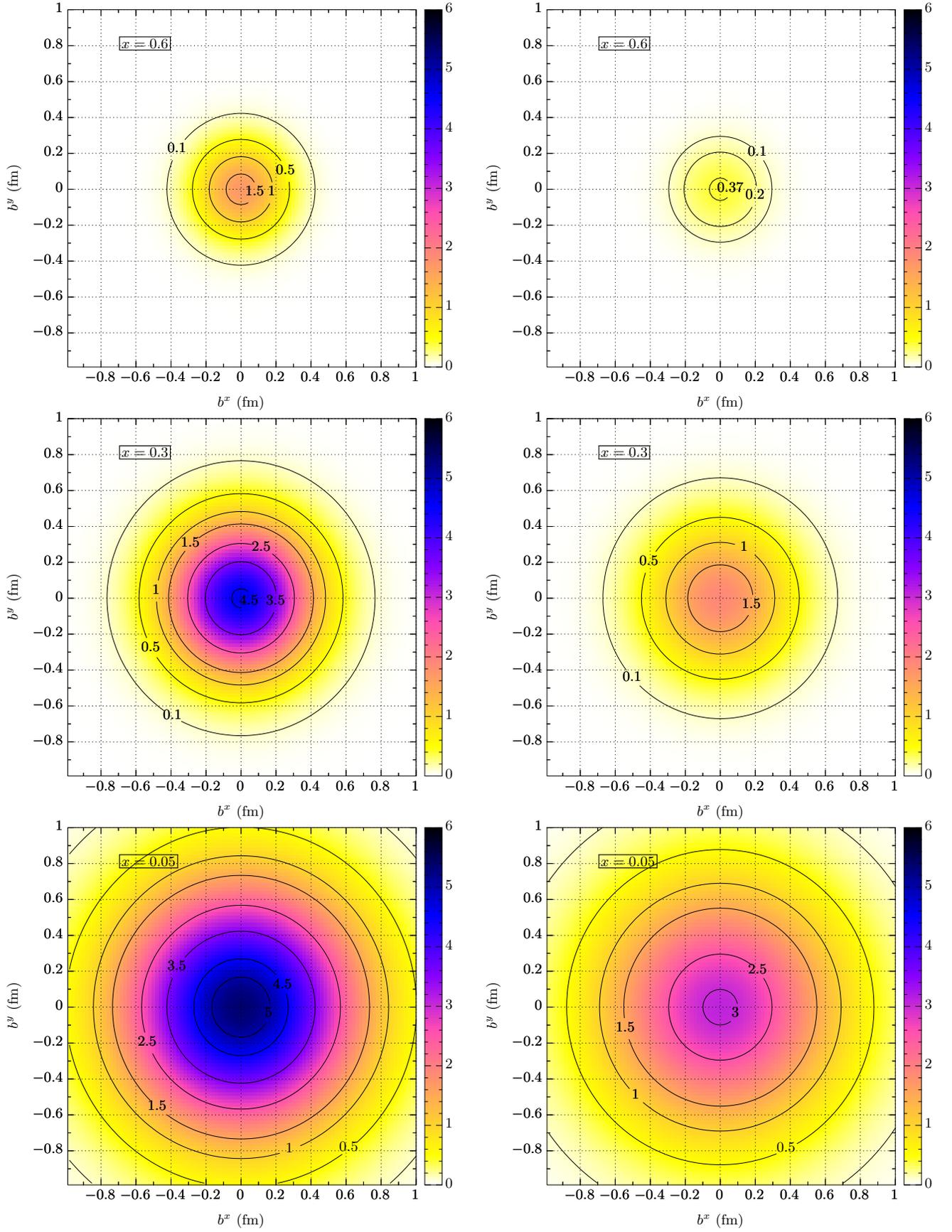

    \centering
    \begin{minipage}{.5\textwidth}
        \centering
\resizebox{!}{0.85\textwidth}
{
\input{./fig-tomo-uv-x06.tex}
}
    \end{minipage}%
    \begin{minipage}{0.5\textwidth}
        \centering
\resizebox{!}{0.85\textwidth}
{
\input{./fig-tomo-dv-x06.tex}
}
    \end{minipage}
        \begin{minipage}{0.5\textwidth}
        \centering
\resizebox{!}{0.85\textwidth}
{
\input{./fig-tomo-uv-x03.tex}
}
    \end{minipage}%
        \begin{minipage}{0.5\textwidth}
        \centering
\resizebox{!}{0.85\textwidth}
{
\input{./fig-tomo-dv-x03.tex}
}
    \end{minipage}
        \begin{minipage}{0.5\textwidth}
        \centering
\resizebox{!}{0.85\textwidth}
{
\input{./fig-tomo-uv-x005.tex}
}
    \end{minipage}%
        \begin{minipage}{0.5\textwidth}
        \centering
\resizebox{!}{0.85\textwidth}
{
\input{./fig-tomo-dv-x005.tex}
}
    \end{minipage}    
    \caption{ Tomography plots of the unpolarized density $ q_v(x,\bm{b}) $ for the up (left frames) and down (right frames) quarks in the transverse $\bm{b}=(b^{x},b^{y})$ plane at three fixed $ x $, namely $ x=0.6, 0.3, 0.05 $. The values of $q_v(x,\mathbf{b})$ have been shown by color and some guiding contours have been drawn. The unit of $q_v(x,\mathbf{b})$ is fm$^{-2}$.}
\label{fig:fig11}
\end{figure*}

\begin{figure*}
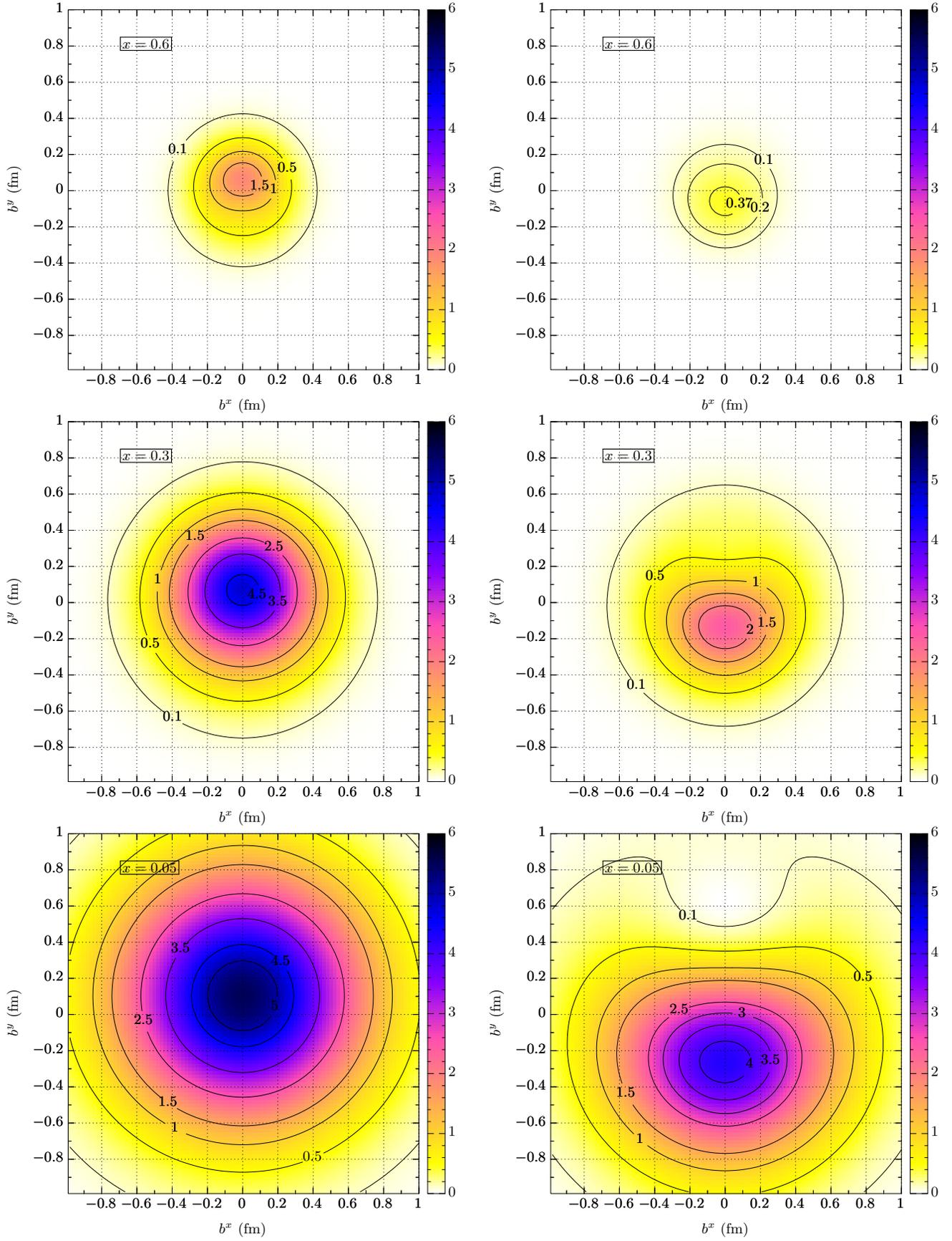

    \centering
    \begin{minipage}{.5\textwidth}
        \centering
\resizebox{!}{0.85\textwidth}
{
\input{./fig-Xtomo-uv-x06.tex}
}
    \end{minipage}%
    \begin{minipage}{0.5\textwidth}
        \centering
\resizebox{!}{0.85\textwidth}
{
\input{./fig-Xtomo-dv-x06.tex}
}
    \end{minipage}
        \begin{minipage}{0.5\textwidth}
        \centering
\resizebox{!}{0.85\textwidth}
{
\input{./fig-Xtomo-uv-x03.tex}
}
    \end{minipage}%
        \begin{minipage}{0.5\textwidth}
        \centering
\resizebox{!}{0.85\textwidth}
{
\input{./fig-Xtomo-dv-x03.tex}
}
    \end{minipage}
        \begin{minipage}{0.5\textwidth}
        \centering
\resizebox{!}{0.85\textwidth}
{
\input{./fig-Xtomo-uv-x005.tex}
}
    \end{minipage}%
        \begin{minipage}{0.5\textwidth}
        \centering
\resizebox{!}{0.85\textwidth}
{
\input{./fig-Xtomo-dv-x005.tex}
}
    \end{minipage}    
    \caption{Same as Fig.~\ref{fig:fig11} but for the density $ q_v^X(x,\bm{b}) $.}
\label{fig:fig12}
\end{figure*}

As a further illustration, in Figs.~\ref{fig:fig11} and~\ref{fig:fig12} we have shown our final results for GPDs as tomography plots in impact parameter space for fixed $ x $. To be more precise, Fig.~\ref{fig:fig11} displays the results of the unpolarized density $ q_v(x,\bm{b}) $ for the up (left frames) and down (right frames) quarks that have been calculated using Eq.~(\ref{Eq:27}) at three fixed values of $ x $, namely $ x=0.6, 0.3, 0.05 $. Figure~\ref{fig:fig12} shows the corresponding results for the unpolarized density $ q_v^X(x,\bm{b}) $ that have been calculated using Eq.~(\ref{Eq:30}) for transversely polarized proton. It is obvious from these figures that the displacement of the center of
the $ q_v^X(x,\bm{b}) $ distribution along the $b^y$-axis is different for the up and down quarks. In fact, it is expected from different signs of $ s_u(x) $ and $ s_d(x) $ that the center of the $ q_v^X(x,\bm{b}) $ density will be shifted toward negative $b^y$ for down quarks, whereas it will be shifted toward positive $b^y$ for up quarks. It should also be noted that the observed difference
between the $ q_v^X(x,\bm{b}) $ and $ q_v(x,\bm{b}) $ for down quark at small $ x $ compared to the corresponding results for the up quark is due to the fact that the shift $ \vert s_d(x) \vert $ is significantly larger than $ s_u(x) $ at small values of $ x $ (see Fig.~\ref{fig:fig10}).


\section{summary and conclusions}\label{sec:five} 
It has been well established that the structure of the nucleon, in both the unpolarized and polarized cases, can be investigated in greater detail using GPDs. According to the DFJK model, GPDs can be expressed in terms of PDFs at zero skewness. In this work, considering DFJK model, we have determined both the unpolarized ($ H $) and polarized ($\widetilde{H}$) GPDs for quarks at NLO using a simultaneous $ \chi^2 $ analysis of the nucleon AFF and WACS experimental data. It can be considered as a continuation of our previous work~\cite{Hashamipour:2019pgy} where we extracted the polarized GPDs, namely HGG19, through a $\chi^2$ analysis of the AFF data solely. The experimental data included in our analysis cover a wide range of the squared transverse momentum $ -t $, i.e., $ 0.025 < -t < 6.46 $ GeV$ ^2 $. In order to find the best set of GPDs, we have performed various analyses of the AFF and WACS data using different approaches. In this regard, we first performed a parameterization scan to find the optimum form of profile function Ansatzes for each flavor of $ H $ and $ \widetilde{H} $ GPDs and reduce the number of free parameters as many as possible. Next, we have considered the three different prescriptions or scenarios for relating the Mandelstam variables at the partonic level to those of the whole process, i.e. at the nucleon level. Then, by performing three separate analyses, we have found the best scenario, namely scenario 3 given by Eq.~(\ref{Eq:21}), for calculating the WACS cross section theoretically which leads to the lowest $ \chi^2 $ and thus the best agreement between the theoretical predictions and experimental data. Moreover, we have performed various analyses using different sets of PDFs to study the sensitivity of the fit results to the PDFs set that we choose for calculating the GPDs $ H $ and the resulting WACS cross section. We have shown that although there are no significant differences between the analyses performed using different sets of PDFs in view of $ \chi^2 $ and parameter values, the extracted GPDs differ as functions of $x$, especially for the case of down valence and sea quark distributions. After finding the optimum form of profile function for each flavor, the best scenario for calculating WACS cross section, and the most compatible PDFs set, we have presented the final results of the extracted GPDs $ H $ and $ \widetilde{H} $ with their uncertainties, namely HGG20, and have compared them with the corresponding ones obtained from other analyses. We have shown that there is a good agreement between the HGG20 and DFJK05~\cite{Diehl:2004cx} unpolarized GPDs, even though the latter has used the experimental data of the Dirac and Pauli FFs, while we have used data from the AFF and WACS. Moreover, we have indicated that the WACS data affect the large $ -t $ behavior of GPDs more considerably. The main result of the present work is that the simultaneous analysis of AFF and WACS data leads to polarized GPDs which differs substantially as compared to the ones obtained by analyzing each of the AFF and WACS data separately (see Fig.~\ref{fig:fig7}). We have compared the theoretical predictions obtained using the final GPDs with the experimental data included in the analysis, and have shown that there is a good agreement in the entire interval of $ -t $ under consideration. Finally, by calculating the distribution in the transverse plane of valence quarks, both in an unpolarized and in a transversely polarized proton, we have presented some tomography plots which illustrate the interplay between longitudinal and transverse partonic degrees of freedom in the proton.
\section*{ACKNOWLEDGMENT}
 H.H. and S.S.G. would like to thank the research council of the Shahid Beheshti University for financial support. M.G. thanks the School of Particles and Accelerators, Institute for Research in Fundamental Sciences (IPM) for financial support provided for this research.

%

\end{document}